\documentclass[pra,aps,superscriptaddress,twocolumn,showpacs]{revtex4}
\usepackage{amsfonts}
\usepackage{amsmath}
\usepackage{amssymb}
\usepackage{mathbbol}
\usepackage{graphicx}
\usepackage{color}

\input{tcilatex}
\begin{document}

\title{Active locking and entanglement in type II optical parametric oscillators}
\author{Joaquin Ruiz-Rivas}
\affiliation{Departament d'\`{O}ptica, Universitat de Val\`{e}ncia, Dr. Moliner 50, 46100--Burjassot, Spain}
\author{Germ\'{a}n J. de Valc\'{a}rcel}
\affiliation{Departament d'\`{O}ptica, Universitat de Val\`{e}ncia, Dr. Moliner 50, 46100--Burjassot, Spain}
\author{Carlos Navarrete-Benlloch}
\affiliation{Institute for Theoretical Physics, Universit\"at Erlangen-N\"urnberg, Staudtstr. 7, 91058 Erlangen, Germany}
\affiliation{Max-Planck-Institut fur Quantenoptik, Hans-Kopfermann-str. 1, 85748 Garching, Germany}
\begin{abstract}
Type II optical parametric oscillators are amongst the highest-quality sources of quantum-correlated light. In particular, when pumped above threshold, such devices generate a pair of bright orthogonally-polarized beams with strong continuous-variable entanglement. However, these sources are of limited practical use, because the entangled beams emerge with different frequencies and a diffusing phase-difference. It has been proven that the use of an internal wave-plate coupling the modes with orthogonal polarization is capable of locking the frequencies of the emerging beams to half the pump frequency, as well as reducing the phase-difference diffusion, at the expense of reducing the entanglement levels. In this work we characterize theoretically an alternative locking mechanism: the injection of a laser at half the pump frequency. Apart from being less invasive, this method should allow for an easier real-time experimental control. We show that such an injection is capable of generating the desired phase locking between the emerging beams, while still allowing for large levels of entanglement. Moreover, we find an additional region of the parameter space (at relatively large injections) where a mode with well defined polarization is in a highly squeezed vacuum state.
\end{abstract}

\maketitle

\section{Introduction}

Optical parametric oscillators (OPOs) are optical cavities
containing a crystal with second order nonlinearity. When pumped with a
laser at frequency $2\omega_{0}$, these are able to generate beams at
frequecies $\omega_{\mathrm{s}}$ (signal) and $\omega_{\mathrm{i}}$ (idler)
such that $\omega_{\mathrm{s}}+\omega_{\mathrm{i}}\approx2\omega_{0}$,
through the nonlinear process known as parametric down-conversion \cite{Boyd03,NavarretePhD}. Classically, the generation of the down-converted field requires
the nonlinear gain to compensate for the cavity losses, what means that the
OPO has to be pumped above a certain threshold power in order for signal and
idler to start oscillating inside the cavity \cite{Boyd03,NavarretePhD}. Quantum
mechanically, on the other hand, down-converted pairs can be generated even
below threshold, what confers the signal and idler fields with very interesting
quantum correlations \cite{Meystre91}.

In particular, type I OPOs, in which both signal and idler are linearly
polarized within the extraordinary axis of the crystal, hold the record for
quadrature noise reduction or single-mode squeezing (more than 90\% below
vacuum fluctuations \cite{Mehmet11,Eberle10,Mehmet10,Vahlbruch08,Takeno07});
this is manifested in the mode at the degenerate frequency $\omega _{\mathrm{%
s}}=\omega_{\mathrm{i}}\sim\omega_{0}$, but squeezing is large only when
working close to threshold \cite{Collett84}. As for the applications of this
quantum-correlated light source, on one hand, squeezed light is a basic
resource in the field of high-precission measurements, helping overcome the
standard quantum limit imposed by vacuum fluctuations \cite{Goda08,Vahlbruch05,Treps02,Treps03}. On the other hand, mixing the output
of two single-mode squeezers on a beam splitter, one can obtain a pair of
entangled beams (in the continuous-variable, Einstein-Podolsky-Rosen sense \cite{EPR}), what makes these devices
a basic resource also for continuous-variable quantum information protocols 
\cite{Braunstein05,Weedbrook12,NavarreteQICVbook}; however, these method for the generation of
entanglement requires the nonlinear cavities to be precisely locked to
generate indistinguishable down-converted fields whose squeezing occurs in
two orthogonal quadratures, which introduces one level of complexity.

Of more interest for our current work are type II OPOs, that is, OPOs in
which signal and idler have orthogonal polarizations (one following the
extraordinary crystal axis, and the other one the ordinary), making the
down-conversion intrinsically non-degenerate \cite{Boyd03,NavarretePhD}. Just as the
degenerate OPO, there is an observable which shows large squeezing levels
only close to threshold, which in this case corresponds to the sum of the
phases of signal and idler; in other words, close to threshold, type II OPOs
show signal-idler phase anticorrelations beyond the standard quantum limit 
\cite{Reid88,Reid89,Drummond90}. But non-degenerate OPOs have one more
interesting property: they are invariant under changes of the signal-idler
phase difference, what means that quantum noise is able to act on this
variable without bounds, making it diffuse and eventually completely
undetermined (in the quantum mechanical sense) \cite{Lane88,Reid88,Reid89bis,Navarrete08,Navarrete10,NavarretePhD}. But, invoking
now the Heisenberg principle, a completely undetermined phase difference
between signal and idler allows for complete noise reduction in their
intensisty difference (its canonically conjugate variable); indeed, signal
and idler become twin beams above threshold, that is, their amplitudes are
perfectly correlated \cite{Reynaud87,Heidmann87,Lane88}. Hence,
non-degenerate OPOs show (ideally) perfect amplitude correlations at any
pumping level above threshold, and large phase anti-correlations close to
threshold, which means that close to this point they should be in a
high-quality continuous-variable entangled state \cite%
{Reid89bis,Reid89,Drummond90,NavarretePhD}. From a quantum optics perspective, this means
that below threshold OPOs should emit a two-mode squeezed vacuum state,
while above threshold OPOs would emit a displaced one (a `bright' EPR state).

However, there are two issues that make above-threshold type II OPOs not
practical as an EPR source, specially from a detection point of view. First,
the phase--matching conditions ensuring that it is the frequency degenerate
process the one with larger gain (lowest threshold) are quite critical, and
hence, signal and idler will have different frequencies in general; for
example, in the case of \cite{Feng03}, where the authors are able to make
the frequency difference between signal and idler as small as 150\textrm{kHz%
} for a cavity with 8\textrm{GHz} free spectral range and 6\textrm{MHz}
linewidth, variations of the cavity length on the order of the nanometer can
make the oscillation frequencies jump to frequencies separated by several
times the free spectral range (mode hopping); second, the signal-idler
phase-difference is chosen at random at any realization and diffuses with
time (rather fast close to threshold), making it virtually impossible to
capture the squeezed quadratures in a balanced homodyne detection scheme.
Hence, additional signal-idler phase locking techniques are required.

The pioneering example of such locking techniques was introduced by Claude Fabre
and collaborators \cite{Longchambon04a,Longchambon04b,Laurat05}. Their idea
consisted in embedding in the cavity a $\lambda/4$ plate with its fast axis
misaligned with respect to the extraordinary axis of the nonlinear crystal.
The plate introduces a coupling between the signal and idler modes which
breaks the phase invariance of the OPO, and it was then shown in \cite%
{Longchambon04a} that in a given region of the parameter space (in
particular of the detunings) the frequencies of signal
and idler get locked to $\omega_{0}$; this OPO is known as the \textit{%
self--phase--locked OPO}, and was already studied experimentally in \cite%
{Laurat05}. Note that, as mentioned, this self--locking effect is
accomplished by breaking the phase symmetry of the OPO, and hence, one
should expect a degradation of the signal-idler intensity correlations, or,
equivalently, of the noncritical squeezing induced by spontaneous
polarization symmetry breaking described in \cite{Garcia10}. For example, in 
\cite{Laurat05} the intensity--difference fluctuations showed 89\% quantum
noise reduction prior to the introduction of the plate, while after
obtaining frequency degeneracy through the self--phase--locking mechanism
this value fell down to a more humble 65\%.

In the present article we study an alternative locking mechanism which
consists in the injection of a laser at frequency degeneracy $\omega_{0}$,
what is less invasive and more controlable at real time than the introduction of a $\lambda/4$ wave plate; we will call \textit{actively--phase--locked OPO} to
such OPO configuration. We show how locking of the signal and idler frequencies to the subharmonic $\omega_0$ can be achieved, while still obtaining large entanglement levels. This locking technique is reminiscent of our previous work in frequency-degenerate type I OPOs tuned to the first
family of transverse modes \cite{Navarrete08,Navarrete10,Garcia09,Navarrete11,Navarrete13,NavarretePhD}, in which we
proposed injecting a TEM$_{10}$ mode at the subharmonic to lock the
phase-difference between the down-converted modes with opposite orbital
angular momentum \cite{Navarrete11}.

The article is organized as follows. In the next section we introduce our
OPO model, providing the set of stochastic equations within the positive P
representation which will allow us to study both its classical and quantum dynamics in
detail. Particularizing to a configuration that we will denote by `symmetric', next we find the classical phase diagram of the system analytically, showing the regimes where frequency locking is expected to appear. Still within this symmetric configuration, we then provide analytical expressions for the quantum correlations of the system, putting special emphasis on the level of signal-idler entanglement at the locking points. In the section before the conclusions, we move out of the symmetric configuration, which is quite challenging to achieve in real experiments, and perform a numerical study that proves all the analytic conclusions of the symmetric case to hold also in this case.

\begin{figure}[t]
\begin{center}
\includegraphics[width=\columnwidth]{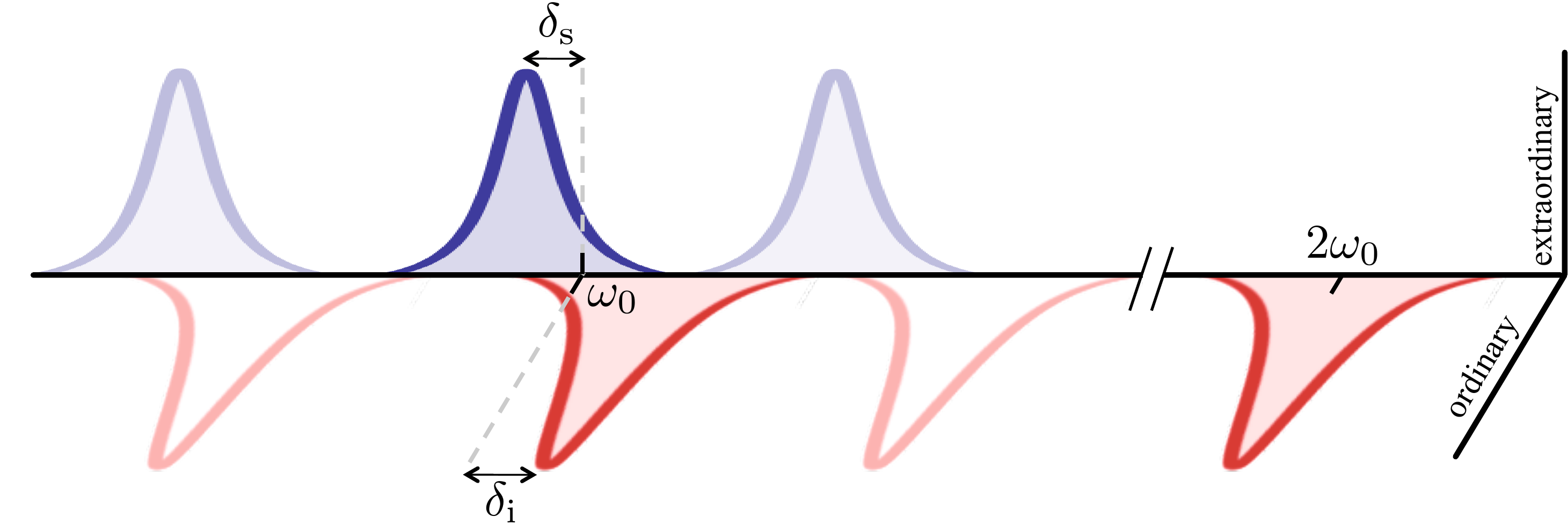}
\end{center}
\caption{Resonance scheme of the type II OPO considered in this work. The birrefringence of the crystal breaks the degeneracy between the modes with ordinary and extraordinary polarization. We show the pump resonance at frequency $2\omega_0$, and three resonances around the subharmonic $\omega_0$, two of which overlap at that frequency and correspond to the signal and idler down-converted modes.}
\label{fTypeII1}
\end{figure}

\section{Model for the actively-phase-locked OPO}

For definiteness and without loss of generality, we consider a symmetric
Fabry-Perot cavity with a thin nonlinear crystal in its center ($z=0$),
where the electric field operator at the relevant frequencies can be
approximately written as $\mathbf{\hat{E}}\left( \mathbf{r}_{\perp},t\right)
=\sum_{j=\mathrm{p},\mathrm{s},\mathrm{i}}\mathbf{\hat{E}}_{j}^{(+)}\left( 
\mathbf{r}_{\perp},t\right) +\mathrm{H.c.}$, with \cite{NavarretePhD}
\begin{equation}
\mathbf{\hat{E}}_{j}^{(+)}\left( \mathbf{r}_{\perp}\mathbf{,}t\right) =%
\mathrm{i}\sqrt{\frac{2\hbar\omega_{j}}{\pi\varepsilon_{0}n_{j}L_{j}w_{j}^{2}%
}}e^{-r^{2}/w_{j}^{2}}\boldsymbol{\varepsilon}_{j}\hat{a}_{j}e^{-(1+\delta_{j%
\mathrm{p}})\mathrm{i}\omega_{0}t}.  \label{Fields}
\end{equation}
The indices `p', `s', and `i' refer to the pump, signal, and idler modes,
respectively. $\omega_{j}$, $n_{j}$, $L_{j}$, $w_{j}$, and $\boldsymbol{%
\varepsilon}_{j}$ are, respectively, the resonance frequency, crystal's
refractive index, optical cavity length, transverse spot size at the cavity
waist, and polarization of the corresponding mode. $\mathbf{r}_{\perp
}=(x,y) $ is the transverse coordinate vector, with $r=|\mathbf{r}_{\perp}|$%
, and we have assumed there are TEM$_{00}$ transverse modes resonating at
the three relevant frequencies, giving rise to the simple Gaussian
transverse profile of the expression. Finally, let us remark that, starting
from the Schrodinger picture, we work in a new picture rotating at frequency 
$2\omega_{0}$ for the pump, and $\omega_{0}$ for signal and idler, so that
the the annihilation ($\hat{a}_{j}$) and creation ($\hat{a}_{j}^{\dagger}$)
operators in the expression are time-independent (only the state of the
system will be time-dependent), and satisfy canonical commutation relations $%
[\hat {a}_{j},\hat{a}_{l}^{\dagger}]=\delta_{jl}$.

The resonance scheme and polarization of the fields are sketched in Fig. \ref{fTypeII1}: the pump is polarized within the ordinary axis of the
crystal and resonates at frequency $2\omega_{0}$, while, by convention, signal and idler are
polarized within the extraordinary and ordinary axis,
respectively, and resonate at frequencies $\omega_{\mathrm{s},\mathrm{i}%
}=\omega_{0}+\delta_{\mathrm{s},\mathrm{i}}$, with $|\delta_{\mathrm{s},%
\mathrm{i}}|$ smaller or on the order of their cavity linewidth $\gamma_{%
\mathrm{s}}=\gamma_{\mathrm{i}}$, taken equal for signal and idler for
simplicity. Apart from pumping the cavity with a laser at frequency $%
2\omega_{0}$ with ordinary polarization, we inject an external laser field
(in phase with the pump drive) at the degenerate frequency $\omega_{0}$ with
polarization $\boldsymbol{\varepsilon}_{0}=e^{-\mathrm{i}\theta_{0}}\mathbf{e%
}_{\mathrm{e}}\cos\varphi_{0}+e^{\mathrm{i}\theta_{0}}\mathbf{e}_{\mathrm{o}%
}\sin \varphi_{0}$, where $\mathbf{e}_{\mathrm{e}}$ and $\mathbf{e}_{\mathrm{%
o}}$ are unit vectors following, respectively, the extraordinary and
ordinary axes of the crystal. Including cavity losses through the usual
Lindblad terms, the master equation governing the evolution of the state of
the system reads \cite{NavarretePhD}
\begin{equation}\label{MasterEq}
\frac{d\hat{\rho}}{dt}=\frac{1}{\mathrm{i}\hbar}\left[ \hat{H},\hat{\rho }%
\right] +\sum_{j=\mathrm{p,s,i}}\gamma_{j}(2\hat{a}_{j}\hat{\rho}\hat{a}%
_{j}^{\dagger}-\hat{a}_{j}^{\dagger}\hat{a}_{j}\hat{\rho}-\hat{\rho}\hat {a}%
_{j}^{\dagger}\hat{a}_{j}),
\end{equation}
in the aforementioned rotating picture where the Hamiltonian can be written
as $\hat{H}=\hat{H}_{0}+\hat{H}_{\mathrm{PDC}}+\hat{H}_{\mathrm{inj}}$, with 
\begin{subequations}
\begin{align}
\hat{H}_{0} & =\hbar\delta_{\mathrm{s}}\hat{a}_{\mathrm{s}}^{\dagger}\hat {a}%
_{\mathrm{s}}+\hbar\delta_{\mathrm{i}}\hat{a}_{\mathrm{i}}^{\dagger}\hat {a}%
_{\mathrm{i}}, \\
\hat{H}_{\mathrm{PDC}} & =\mathrm{i}\hbar\chi(\hat{a}_{\mathrm{p}}\hat {a}_{%
\mathrm{s}}^{\dagger}\hat{a}_{\mathrm{i}}^{\dagger}-\hat{a}_{\mathrm{p}%
}^{\dagger}\hat{a}_{\mathrm{s}}\hat{a}_{\mathrm{i}}), \\
\hat{H}_{\mathrm{inj}} & =\sum_{j=\mathrm{p,s,i}}\mathrm{i}\hbar (\mathcal{E}%
_{j}\hat{a}_{j}^{\dagger}-\mathcal{E}_{j}^{\ast}\hat{a}_{j}).
\end{align}
In this expression, the down-conversion rate $\chi$ is
proportional to the crystal's nonlinear susceptibility, and the damping
rates are related to the (intensity) transmisivities of the mirror at the
corresponding frequency, $\mathcal{T}_{j}$, by $\gamma_{j}\approx c\mathcal{T%
}_{j}/4L_{j}$. In addition, the injection parameters can be approximately
written in terms of the power $P_{j}$ of the injected lasers at frequencies $%
2\omega_{0}$ and $\omega_{0}$ as $\mathcal{E}_{\mathrm{p}}=\sqrt{\gamma_{%
\mathrm{p}}P_{2\omega_{0}}/\hbar\omega_{0}}$, $\mathcal{E}_{\mathrm{s}}=%
\sqrt {2\gamma_{\mathrm{s}}P_{\omega_{0}}/\hbar\omega_{0}}e^{-\mathrm{i}%
\theta_{0}}\cos\varphi_{0}$, and $\mathcal{E}_{\mathrm{i}}=\sqrt{2\gamma_{%
\mathrm{s}}P_{\omega_{0}}/\hbar\omega_{0}}e^{\mathrm{i}\theta_{0}}\sin%
\varphi_{0}$, where we have taken the phase of the driving lasers as a
reference.

In order to get analytical insight, and following previous works \cite{Navarrete08,Navarrete10,Navarrete11,Navarrete13}, we
map this master equation into a set of stochastic Langevin equations by
using the positive $P$ coherent representation \cite{Drummond80}. This is an exact
procedure by which an independent complex stochastic variable is associated
to each bosonic operator, that is, $\{\alpha _{j},\alpha _{j}^{+}\}_{j=%
\mathrm{p,s,i}}$ to $\{\hat{a}_{j},\hat{a}_{j}^{\dagger }\}_{j=\mathrm{p,s,i}%
}$; quantum expectation values of any operator are then obtained as
stochastic averages by replacing the bosonic operators by their
corresponding stochastic variable in the normally-ordered version of the
operator. It is not difficult to show \cite{NavarretePhD} that the stochastic Langevin
equations associated to the master equation (\ref{MasterEq}) read 
\end{subequations}
\begin{subequations}
\begin{align}
\dot{\alpha}_{\mathrm{p}}& =\mathcal{E}_{\mathrm{p}}-\gamma _{\mathrm{p}%
}\alpha _{\mathrm{p}}-\chi \alpha _{\mathrm{s}}\alpha _{\mathrm{i}}, \\
\dot{\alpha}_{\mathrm{p}}^{+}& =\mathcal{E}_{\mathrm{p}}-\gamma _{\mathrm{p}%
}\alpha _{\mathrm{p}}^{+}-\chi \alpha _{\mathrm{s}}^{+}\alpha _{\mathrm{i}%
}^{+}, \\
\dot{\alpha}_{\mathrm{s}}& =\mathcal{E}_{\mathrm{s}}-\left( \gamma _{\mathrm{%
s}}+\mathrm{i}\delta _{\mathrm{s}}\right) \alpha _{\mathrm{s}}+\chi \alpha _{%
\mathrm{p}}\alpha _{\mathrm{i}}^{+}+\sqrt{\chi \alpha _{\mathrm{p}}}\xi (t),
\\
\dot{\alpha}_{\mathrm{s}}^{+}& =\mathcal{E}_{\mathrm{s}}-\left( \gamma _{%
\mathrm{s}}-\mathrm{i}\delta _{\mathrm{s}}\right) \alpha _{\mathrm{s}%
}^{+}+\chi \alpha _{\mathrm{p}}^{+}\alpha _{\mathrm{i}}+\sqrt{\chi \alpha _{%
\mathrm{p}}^{+}}\xi ^{+}(t), \\
\dot{\alpha}_{\mathrm{i}}& =\mathcal{E}_{\mathrm{i}}-\left( \gamma _{\mathrm{%
s}}+\mathrm{i}\delta _{\mathrm{i}}\right) \alpha _{\mathrm{i}}+\chi \alpha _{%
\mathrm{p}}\alpha _{\mathrm{s}}^{+}+\sqrt{\chi \alpha _{\mathrm{p}}}\xi
^{\ast }(t), \\
\dot{\alpha}_{\mathrm{i}}^{+}& =\mathcal{E}_{\mathrm{i}}-\left( \gamma _{%
\mathrm{s}}-\mathrm{i}\delta _{\mathrm{i}}\right) \alpha _{\mathrm{i}%
}^{+}+\chi \alpha _{\mathrm{p}}^{+}\alpha _{\mathrm{s}}+\sqrt{\chi \alpha _{%
\mathrm{p}}^{+}}\xi ^{+\ast }(t),
\end{align}%
where we have defined independent complex noises $\xi (t)$ and $\xi ^{+}(t)$%
, with zero mean, and only non-zero two-time correlators 
\end{subequations}
\begin{equation}\label{ComplexNoiseCorr}
\left\langle \xi (t)\xi ^{\ast }(t^{\prime })\right\rangle =\left\langle \xi
^{+}(t)\xi ^{+\ast }(t^{\prime })\right\rangle =\delta (t-t^{\prime }).
\end{equation}

In order to reduce the number of parameters of the problem, we now make some
variable changes; in particular, we redefine time as $\tau =\gamma _{\mathrm{%
s}}t$, the coherent amplitudes as 
\begin{equation}
\beta _{\mathrm{p}}=\frac{\chi }{\gamma _{\mathrm{s}}}\alpha _{\mathrm{p}},%
\text{ \ \ }\beta _{\mathrm{s,i}}=\frac{\chi }{\sqrt{\gamma _{\mathrm{s}%
}\gamma _{\mathrm{p}}}}\alpha _{\mathrm{s,i}}\exp (\pm \mathrm{i}\theta
_{0}),
\end{equation}%
and the noises as%
\begin{equation}
\eta (\tau )=\frac{1}{\sqrt{\gamma _{\mathrm{s}}}}\xi (t),\text{ \ \ }\eta
^{+}(\tau )=\frac{1}{\sqrt{\gamma _{\mathrm{s}}}}\xi ^{+}(t),
\end{equation}%
which satisfy the statistical properties (\ref{ComplexNoiseCorr}), but now
respect to the dimensionless time $\tau $. In terms of these new variables,
the Langevin equations read 
\begin{subequations}
\label{NormLangevin}
\begin{align}
\dot{\beta}_{\mathrm{p}}& =\kappa \left( \sigma -\beta _{\mathrm{p}}-\beta _{%
\mathrm{s}}\beta _{\mathrm{i}}\right) , \\
\dot{\beta}_{\mathrm{p}}& =\kappa \left( \sigma -\beta _{\mathrm{p}%
}^{+}-\beta _{\mathrm{s}}^{+}\beta _{\mathrm{i}}^{+}\right) , \\
\dot{\beta}_{\mathrm{s}}& =\varepsilon _{\mathrm{s}}-\left( 1+\mathrm{i}%
\Delta _{\mathrm{s}}\right) \beta _{\mathrm{s}}+\beta _{\mathrm{p}}\beta _{%
\mathrm{i}}^{+} \\
& \text{ \ \ \ \ \ \ \ \ \ \ \ \ \ \ \ \ \ \ }+g\sqrt{\beta _{\mathrm{p}}}%
\exp (\mathrm{i}\theta _{0})\eta (\tau ),  \notag \\
\dot{\beta}_{\mathrm{s}}^{+}& =\varepsilon _{\mathrm{s}}-\left( 1-\mathrm{i}%
\Delta _{\mathrm{s}}\right) \beta _{\mathrm{s}}^{+}+\beta _{\mathrm{p}%
}^{+}\beta _{\mathrm{i}} \\
& \text{ \ \ \ \ \ \ \ \ \ \ \ \ \ \ \ \ \ \ }+g\sqrt{\beta _{\mathrm{p}}^{+}%
}\exp (-\mathrm{i}\theta _{0})\eta ^{+}(\tau ),  \notag \\
\dot{\beta}_{\mathrm{i}}& =\varepsilon _{\mathrm{i}}-\left( 1+\mathrm{i}%
\Delta _{\mathrm{i}}\right) \beta _{\mathrm{i}}+\beta _{\mathrm{p}}\beta _{%
\mathrm{s}}^{+} \\
& \text{ \ \ \ \ \ \ \ \ \ \ \ \ \ \ \ \ \ \ }+g\sqrt{\beta _{\mathrm{p}}}%
\exp (-\mathrm{i}\theta _{0})\eta ^{\ast }(\tau ),  \notag \\
\dot{\beta}_{\mathrm{i}}& =\varepsilon _{\mathrm{i}}-\left( 1-\mathrm{i}%
\Delta _{\mathrm{i}}\right) \beta _{\mathrm{i}}^{+}+\beta _{\mathrm{p}%
}^{+}\beta _{\mathrm{s}} \\
& \text{ \ \ \ \ \ \ \ \ \ \ \ \ \ \ \ \ \ \ }+g\sqrt{\beta _{\mathrm{p}}^{+}%
}\exp (\mathrm{i}\theta _{0})\eta ^{+\ast }(\tau ),  \notag
\end{align}%
where we have defined the parameters 
\end{subequations}
\begin{align}
\kappa & =\frac{\gamma _{\mathrm{p}}}{\gamma _{\mathrm{s}}},\text{\ \ \ }%
\sigma =\frac{\chi \mathcal{E}_{\mathrm{p}}}{\gamma _{\mathrm{s}}\gamma _{%
\mathrm{p}}},\text{\ \ \ }\Delta _{j}=\frac{\delta_{j}}{\gamma_{\mathrm{s}}}, \\
\varepsilon _{\mathrm{s,i}}& =\frac{g}{\gamma _{\mathrm{s}}}|\mathcal{E}_{%
\mathrm{s,i}}|\text{,\ \ \ }g=\frac{\chi }{\sqrt{\gamma _{\mathrm{s}}\gamma
_{\mathrm{p}}}}\text{.}  \notag
\end{align}%
Note that the Fokker-Planck equation associated to this Langevin system is
independent of $\theta _{0}$, and hence, we can ignore the phase factors in
the noises. In other words, the system is only sensitive to the parameter $%
\varphi _{0}$ of the injection's polarization.

\begin{figure*}[t]
\begin{center}
\includegraphics[width=0.75\textwidth]{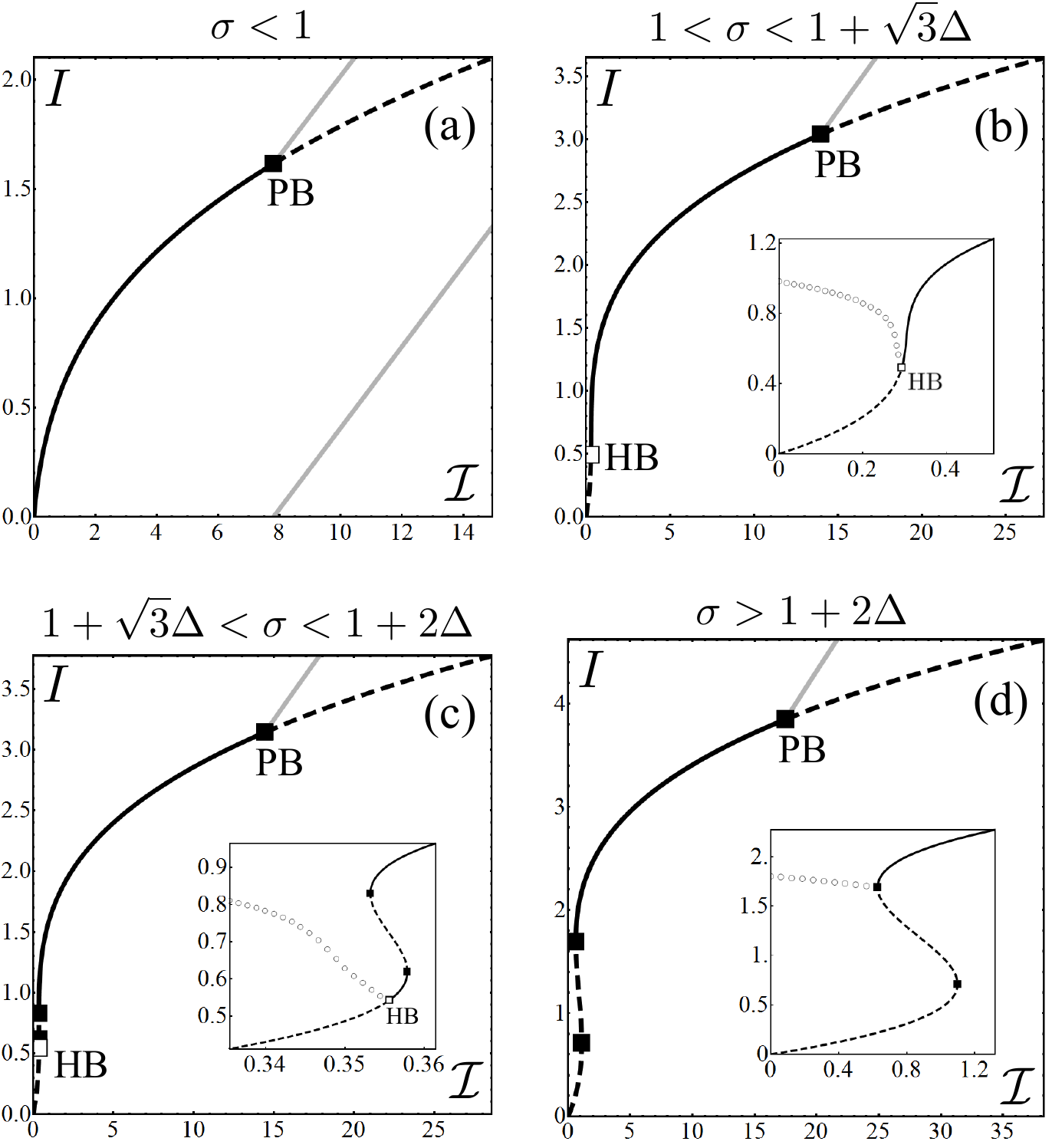}
\end{center}
\caption{Bifurcation diagrams of the system. The value $\Delta=0.6$ is chosen for all the figures (the same behavior is found for any other choice), while we set $\sigma$ to 0.5 in (a), 1.98 in (b), 2.09 in (c), and 2.8 in (d). The black lines correspond to the intensity $I$ of the stationary symmetric solution (\ref{SymmSol}), their solid or dashed character meaning that this solution is stable or unstable, respectively. The upper and lower grey solid lines correspond to the values of $|\bar{\beta}_{\mathrm{b}}|^{2}/2$ and $|\bar{\beta}_{\mathrm{d}}|^{2}/2$, respectively, that is to half the intensity of the bright and dark modes (the latter only showed in the $\sigma<1$ case); these lines have been found numerically, and show how above the pitchfork bifurcation (marked as PB in the
figures) the symmetric solution (\ref{SymmSol}) becomes unstable, and a new asymmetric solution is born with $|\bar{\beta}_\text{s}|\neq|\bar{\beta}_\text{i}|$. As explained in the text, for $\sigma>1$ it is
possible to find periodic solutions connecting the $\mathcal{I}=0$ axis with the Hopf bifurcation (marked as HB in the figures); we have checked numerically that this periodic orbits exist, and moreover they are `symmetric', that is, $\beta_{\mathrm{s}}(\tau)=\beta_{\mathrm{i}}^{\ast}(\tau)$. The grey circles correspond in this case to the mean value of $|\beta_{\mathrm{s}}(\tau)|^{2}$ (half the sum between its maximum and its minimum of oscillation). Note that there exist regions where stable stationary
solutions and periodic orbits coexist, and that after the Hopf bifurcation is extinguished ($\sigma>1+2\Delta$) the periodic orbits are connected
directly to the upper turning point of the S--shaped curve.}
\label{fTypeII2}
\end{figure*}

In order to get some analytic insight in the rest of the article (with the exception of the last section) we are going to simplify the problem
to what we will call \textit{symmetric configuration} of the
actively--phase--locked OPO: We assume the detunings to be opposite, that
is, $\Delta _{\mathrm{s}}=-\Delta _{\mathrm{i}}=\Delta >0$, and inject with $%
\varphi _{0}=\pi /4$ (arbitrary polarization ellipse along the $\pm 45%
{{}^o}%
$ axis), so that signal and idler get equally pumped, $|\varepsilon _{%
\mathrm{s}}|=|\varepsilon _{\mathrm{i}}|\equiv \sqrt{\mathcal{I}}$.
Furthermore, we consider the $\kappa \gg 1$ limit in which the pump can be
adiabatically eliminated ($\dot{\beta}_{\mathrm{p}}=\dot{\beta}_{\mathrm{p}%
}^{+}=0$ in the previous equations). Taking all these considerations into
account, we can reduce our model equations (\ref{NormLangevin}) to 
\begin{subequations}
\label{FinalLangevin}
\begin{align}
\dot{\beta}_{\mathrm{s}}& =\sqrt{\mathcal{I}}-\left( 1+\mathrm{i}\Delta
\right) \beta _{\mathrm{s}}+\tilde{\beta}_{\mathrm{p}}\beta _{\mathrm{i}}^{+}+g%
\sqrt{\tilde{\beta}_{\mathrm{p}}}\eta (\tau ),  \label{NormEqs1} \\
\dot{\beta}_{\mathrm{s}}^{+}& =\sqrt{\mathcal{I}}-\left( 1-\mathrm{i}\Delta
\right) \beta _{\mathrm{s}}^{+}+\tilde{\beta}_{\mathrm{p}}^{+}\beta _{\mathrm{i%
}}+g\sqrt{\tilde{\beta}_{\mathrm{p}}^{+}}\eta ^{+}(\tau ),  \label{NormEqs2} \\
\dot{\beta}_{\mathrm{i}}& =\sqrt{\mathcal{I}}-\left( 1-\mathrm{i}\Delta
\right) \beta _{\mathrm{i}}+\tilde{\beta}_{\mathrm{p}}\beta _{\mathrm{s}}^{+}+g%
\sqrt{\tilde{\beta}_{\mathrm{p}}}\eta ^{\ast }(\tau ),  \label{NormEqs3} \\
\dot{\beta}_{\mathrm{i}}^{+}& =\sqrt{\mathcal{I}}-\left( 1+\mathrm{i}\Delta
\right) \beta _{\mathrm{i}}^{+}+\tilde{\beta}_{\mathrm{p}}^{+}\beta _{\mathrm{s%
}}+g\sqrt{\tilde{\beta}_{\mathrm{p}}^{+}}\eta ^{+\ast }(\tau ),
\label{NormEqs4}
\end{align}%
with $\tilde{\beta}_{\mathrm{p}}=\sigma -\beta _{\mathrm{s}}\beta _{\mathrm{i}}
$ and $\tilde{\beta}_{\mathrm{p}}^{+}=\sigma -\beta _{\mathrm{s}}^{+}\beta
_{\mathrm{i}}^{+}.$

These are the final equations that will model quantum-mechanically our
system in the remaining of the paper. In this work we are interested in the
quantum properties of the down-converted field. In particular, defining a
polarization mode
\end{subequations}
\begin{equation}
\boldsymbol{\varepsilon}_{\theta}=[e^{-\mathrm{i}\left( \theta_{\mathrm{0}%
}-\theta\right) }\mathbf{e}_{\mathrm{e}}+e^{\mathrm{i}\left( \theta _{%
\mathrm{0}}-\theta\right) }\mathbf{e}_{\mathrm{o}}]/\sqrt{2},
\end{equation}
where we include in the definition the phase $\theta_{0}$ of the injection
beam for later convenience, with associated annihilation operator%
\begin{equation}
\hat{a}_{\theta}=[e^{\mathrm{i}\left( \theta_{0}-\theta\right) }\hat {a}_{%
\mathrm{s}}+e^{-\mathrm{i}\left( \theta_{0}-\theta\right) }\hat {a}_{\mathrm{%
i}}]/\sqrt{2},
\end{equation}
we will be interested in the noise spectrum associated to one of its
quadratures $\hat{X}_{\theta}^{\psi}=e^{-\mathrm{i}\psi}\hat{a}_{\theta }+e^{%
\mathrm{i}\psi}\hat{a}_{\theta}^{\dagger}$, which can be obtained as

\begin{eqnarray}\label{Vout}
&&V^{\mathrm{out}}(\hat{X}_{\theta }^{\psi };\Omega )=1 \\
&&\text{ \ \ \ \ \ \ \ }+\frac{2}{g^{2}}\lim_{\tau \rightarrow \infty
}\int_{-\infty }^{+\infty }d\tau ^{\prime }\langle \delta x_{\theta }^{\psi
}(\tau )\delta x_{\theta }^{\psi }(\tau +\tau ^{\prime })\rangle e^{-\mathrm{%
i}\Omega \tau ^{\prime }},  \notag
\end{eqnarray}%
where $\Omega $ is the so-called noise frequency (normalized to $\gamma _{%
\mathrm{s}}$), and $x_{\theta }^{\psi }=e^{-\mathrm{i}\psi }\beta _{\theta
}+e^{\mathrm{i}\psi }\beta _{\theta }^{+}$ is the stochastic variable
accounting for the quadrature associated to the normalized stochastic
amplitudes
\begin{subequations}
\begin{align}
\beta _{\theta }& =(e^{-\mathrm{i}\theta }\beta _{\mathrm{s}}+e^{\mathrm{i}%
\theta }\beta _{\mathrm{i}})/\sqrt{2}, \\
\beta _{\theta }^{+}& =(e^{\mathrm{i}\theta }\beta _{\mathrm{s}}^{+}+e^{-%
\mathrm{i}\theta }\beta _{\mathrm{i}}^{+})/\sqrt{2}.
\end{align}%
We have also introduced the notation $\delta x_{\theta }^{\psi }=x_{\theta
}^{\psi }-\langle x_{\theta }^{\psi }\rangle $. This noise spectrum is the
quantity usually measured in a homodyne detection of the field coming out of
the cavity when the local oscillator matches the spatio-temporal profile of
the down-converted field, and has polarization $\boldsymbol{\varepsilon }%
_{\theta }$ and phase $\psi $ relative to the pump. Quantum correlations are
manifest whenever $V(\hat{X}_{\theta }^{\psi };\Omega )<1$ for some value of
the parameters, in which case we say that quadrature $\hat{X}_{\theta
}^{\psi }$ is squeezed at noise frequency $\Omega $. Let us finally remark
that in the following we will use the notation $\hat{Y}_{\theta }^{\varphi }=%
\hat{X}_{\theta }^{\varphi +\pi /2}$\ and\ $y_{\theta }^{\varphi }=x_{\theta
}^{\varphi +\pi /2}$ when needed.

\section{Classical behavior: frequency locking}

Let us first analyze the classical behavior of the system, which will allow us
to see the regions of the parameter space where the signal and idler
oscillation frequencies get locked. The classical limit can be retrieved by making a coherent-state ansatz for all fields, whose amplitude plays the role of the (normalized) amplitude of the classical electromagnetic fields. Within the positive \textit{P} representation, this is equivalent to replacing the `plus' amplitudes by the corresponding complex-conjugate ones and setting the noises to zero, leading to 
\end{subequations}
\begin{subequations}
\label{InjOPOeqsNorm}
\begin{align}
\dot{\beta}_{\mathrm{s}} & =\sqrt{\mathcal{I}}-\left( 1+\mathrm{i}%
\Delta\right) \beta_{\mathrm{s}}+(\sigma-\beta_{\mathrm{s}}\beta_{\mathrm{i}%
})\beta_{\mathrm{i}}^{\ast}, \\
\dot{\beta}_{\mathrm{i}} & =\sqrt{\mathcal{I}}-\left( 1-\mathrm{i}%
\Delta\right) \beta_{\mathrm{i}}+(\sigma-\beta_{\mathrm{s}}\beta_{\mathrm{i}%
})\beta_{\mathrm{s}}^{\ast}.
\end{align}
The oscillation frequency of the classical fields $\mathbf{\hat{E}}%
_{\mathrm{s,i}}^{(+)}\left( \mathbf{r}_{\perp}\mathbf{,}t\right)\vert
_{\hat{a}_{j}\rightarrow\alpha_{j}}$ will be locked to $\omega _{0}$
whenever this nonlinear system has a stationary solution, see Eq. (\ref%
{Fields}). On the other hand, the symmetry $\{\beta_{\mathrm{s}%
}\rightarrow\beta_{\mathrm{i}}^{\ast},\beta_{\mathrm{i}}\rightarrow \beta_{%
\mathrm{s}}^{\ast}\}$ of these equations suggests looking for stationary
solutions of the type 
\end{subequations}
\begin{equation}
\bar{\beta}_{\mathrm{s}}=\bar{\beta}_{\mathrm{i}}^{\ast}=\sqrt{I}\exp(%
\mathrm{i}\varphi).  \label{SymmSol}
\end{equation}
In the remaining of this section we study the conditions under which this
type of solutions exist and are stable.

First, it is straightforward to show from (\ref{InjOPOeqsNorm}) that the
intensity $I$ of this \textit{symmetric solution} satisfies the third order polynomial%
\begin{equation}
\mathcal{I}=[(I+1-\sigma)^{2}+\Delta^{2}]I,  \label{Ipol}
\end{equation}
with its phase $\varphi$ uniquely determined from $I$ as $\varphi
=\arg\{I+1-\sigma-\mathrm{i}\Delta\}$. Depending on the parameters, this
polynomial can have one or three positive definite solutions (see Fig. \ref%
{fTypeII2}); by solving the equation $\partial\mathcal{I}/\partial I=0$, it
is simple to show that the turning points $I_{\pm}$ have the expression%
\begin{equation}
I_{\pm}=\frac{2}{3}(\sigma-1)\pm\frac{1}{3}\sqrt{(\sigma-1)^{2}-3\Delta^{2}},
\end{equation}
and hence, they exist only for $\sigma>1+\sqrt{3}\Delta$. For $\sigma \leq1+%
\sqrt{3}\Delta$~the solution is therefore single-valued.

In order to analyze the stability of this symmetric solution, we will change
to a new polarization basis $\boldsymbol{\varepsilon }_{\mathrm{b}}=%
\boldsymbol{\varepsilon }_{\varphi }$ and $\boldsymbol{\varepsilon }_{%
\mathrm{d}}=\boldsymbol{\varepsilon }_{\varphi -\pi /2},$ where $\boldsymbol{%
\varepsilon }_{\mathrm{b}}$ corresponds to the polarization mode excited by
the symmetric solution (\ref{SymmSol}) and $\boldsymbol{\varepsilon }_{%
\mathrm{d}}$ to its orthogonal, that is, to what we will call the \textit{%
bright} and \textit{dark} modes of the system, as we did in previous works
\cite{Navarrete08,Navarrete10,Navarrete11,Navarrete13,Garcia09,Garcia10,NavarretePhD}. The corresponding normalized amplitudes satisfy the evolution
equations 
\begin{subequations}
\begin{align}
\dot{\beta}_{\mathrm{b}}& =\sqrt{2\mathcal{I}}\cos \varphi -\beta _{\mathrm{b%
}}-\Delta \beta _{\mathrm{d}}+\left( \sigma -\frac{\beta _{\mathrm{b}}^{2}}{2%
}-\frac{\beta _{\mathrm{d}}^{2}}{2}\right) \beta _{\mathrm{b}}^{\ast }, \\
\dot{\beta}_{\mathrm{d}}& =\sqrt{2\mathcal{I}}\sin \varphi -\beta _{\mathrm{d%
}}+\Delta \beta _{\mathrm{b}}+\left( \sigma -\frac{\beta _{\mathrm{b}}^{2}}{%
2}-\frac{\beta _{\mathrm{d}}^{2}}{2}\right) \beta _{\mathrm{d}}^{\ast }.
\end{align}%
In this new basis the symmetric solution (\ref{SymmSol}) simply reads $\{%
\bar{\beta}_{\mathrm{b}}=\sqrt{2I},$ $\bar{\beta}_{\mathrm{d}}=0\}$ and its
associated stability matrix is 
\end{subequations}
\begin{equation}
\mathcal{L}=%
\begin{bmatrix}
-1-2I & \sigma -I & -\Delta  & 0 \\ 
\sigma -I & -1-2I & 0 & -\Delta  \\ 
\Delta  & 0 & -1 & \sigma -I \\ 
0 & \Delta  & \sigma -I & -1%
\end{bmatrix}%
.  \label{Lbd}
\end{equation}%
The characteristic polynomial of this stability matrix can be factorized
into two second order polynomials, namely $P_{\mathrm{I}}(\lambda )=(\lambda
+1+\sigma )^{2}+\Delta ^{2}-I^{2}$ and $P_{\mathrm{II}}(\lambda )=(\lambda
+1-\sigma +2I)^{2}+\Delta ^{2}-I^{2}$. The bifurcation diagrams for the
different parameter regions are shown in Fig. \ref{fTypeII2}; now we
discuss them in length.

Let us start by studying the instabilities predicted by the first
polynomial, whose roots are given by%
\begin{equation}
\lambda_{\pm}^{\mathrm{I}}=-(1+\sigma)\pm\sqrt{I^{2}-\Delta^{2}}.
\end{equation}
The condition $\text{Re}\{\lambda_{\pm}^{\mathrm{I}}\}=0$ can only be
satisfied for%
\begin{equation}
I=\sqrt{(1+\sigma)^{2}+\Delta^{2}}\equiv I_{\mathrm{PB}}\text{.}
\end{equation}
The fact that the instability appears without imaginary part in $%
\lambda_{\pm}^{\mathrm{I}}$, and it is located in the upper branch of the
S--shaped curve ($I_{\mathrm{PB}}>I_{+}$ for any value of the parameters),
signals that it corresponds to a Pitchfork or static bifurcation where an asymmetric
stationary solution with $|\bar{\beta}_\text{s}|\neq|\bar{\beta}_\text{i}|$ borns (as we have checked numerically, see the grey lines in
Fig. \ref{fTypeII2}). This bifurcation is similar to the one introduced in 
\cite{Navarrete11}, where we studied the effects of a signal injection in the
two-transverse-mode DOPO, and can be understood as a switching on of the
dark mode. However, note that in this case the fluctuations of the bright
and dark modes are not decoupled below threshold, see the linear stability
matrix (\ref{Lbd}), what physically means that the quantum properties of
the dark mode at the bifurcation will be different from those of the dark
mode in \cite{Navarrete11}, and hence no perfect squeezing is likely to be
found, as we show later.

As for the second polynomial, its roots are given by%
\begin{equation}
\lambda_{\pm}^{\mathrm{II}}=\sigma-1-2I\pm\sqrt{I^{2}-\Delta^{2}}.
\end{equation}
Note that $\lambda_{\pm}^{\mathrm{II}}=0$ for $I=I_{\pm}$, that is, the
turning points of the S--shaped curve correspond to bifurcation points. It
is then simple to check (for example numerically) that the whole middle
branch connecting this instability points is unstable, a characteristic trade of intensity-bistable systems
(see Figs. \ref{fTypeII2}c,d).

But $\lambda_{\pm}^{\mathrm{II}}$ has yet one more instability when
\begin{equation}
I=\frac{\sigma-1}{2}\equiv I_{\mathrm{HB}},
\end{equation}
provided $1<\sigma<1+2\Delta$. At this instability the eigenvalues become purely imaginary, in particular, $\lambda_{\pm}^{\mathrm{II}}=\pm\mathrm{i}\omega_{\mathrm{HB}}$ with $\omega_{\mathrm{HB}}=\sqrt{\Delta^{2}-(\sigma-1)^{2}/4}$, and hence it corresponds
to a Hopf bifurcation. It is simple to check that $I_{\mathrm{HB}}$ is always below $I_{\mathrm{PB%
}}$ and $I_{-}$; in particular, it borns at $I=0$ for $\sigma=1$, and climbs
the $\mathcal{I}-I$ curve as $\sigma$ increases until it dies at $I=I_{-}$
for $\sigma=1+2\Delta$ (see Figs. \ref{fTypeII2}b,c,d). The portion of the
curve with $I<I_{\mathrm{HB}}$ is unstable, and no stationary solutions can
be found there, as the stable states correspond in this case to periodic
orbits (as we have checked numerically, see Figs. \ref{fTypeII2}b,c,d).
This is also quite intuitive because, when no injection is present ($\mathcal{I}=0$), we know that the stable states of the OPO above
threshold are the ones with the signal and idler beams oscillating at the
non-degenerate frequencies $\omega_{\mathrm{s}}=\omega _{0}+\gamma_{\mathrm{s%
}}\Delta$ and $\omega_{\mathrm{i}}=\omega_{0}-\gamma_{\mathrm{s}}\Delta$,
which in the picture we are working on means $\beta_{\mathrm{s}}(\tau)\propto\exp(-\mathrm{i}\Delta\tau)$ and $\beta _{\mathrm{i}}(\tau)\propto\exp(\mathrm{i}\Delta\tau)$.

This analysis proves that there exist regions in the parameter space where
the frequencies of the signal and idler beams are locked to the degenerate
one, and hence active--locking can be a good alternative to the self-locking
technique already proposed for type II OPOs \cite{Longchambon04a,Longchambon04b,Laurat05}.

\begin{figure*}[t]
\begin{center}
\includegraphics[width=\textwidth]{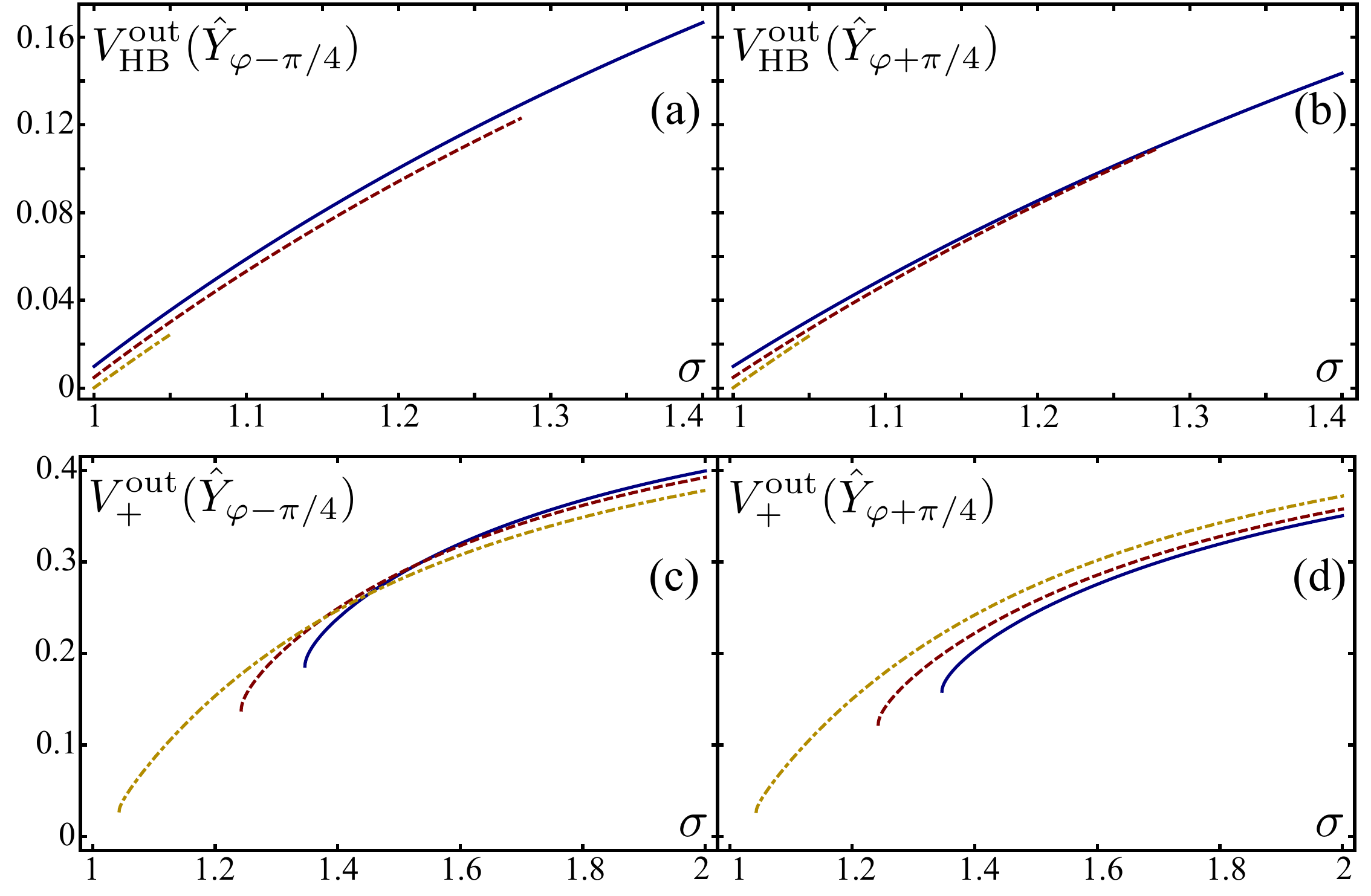}
\end{center}
\caption{Zero-frequency noise spectra of the $\hat{Y}$ quadrature of the $\boldsymbol{\varepsilon}_{\varphi\pm\pi/4}$ modes at the Hopf bifurcation
(up) and the upper turning point (down), as a function of the pump parameter $\sigma$. Three values of $\Delta$ are considered: $0.2$ (solid blue), $0.14$ (dashed red), and $0.025$ (dotted-dashed yellow), the last one corresponding to the value obtained in \cite{Feng03}.}
\label{fTypeII3}
\end{figure*}

\section{Quantum properties}

As explained in the introduction, in the absence of subharmonic injection ($%
\mathcal{I}=0$), it is well known that there is perfect entanglement between
the signal and idler modes for $\sigma=1$ within the linearized description; above this threshold, the
entanglement level is degraded (although perfect amplitude correlations
persist), and the signal and idler fields start oscillating at different
frequencies. Our main intention with the injection was to lock these frequencies to the degenerate one, $\omega_0$,
which should contribute to make the observation and use of their entanglement simpler, since we will
show the entanglement to be equivalent to squeezing in a couple of modes
with well defined frequency and orthogonal polarization. We expect
the presence of the injection to degrade the entanglement level, since it
breaks the phase invariance of the OPO, and in this section we are going to
evaluate how fragile the entanglement is to this injection, proving that large entanglement can still be attained.

In order to analyze the quantum properties of the system, let us first move
again to the basis defined by the bright and dark modes, $\boldsymbol{%
\varepsilon }_{\mathrm{b}}=\boldsymbol{\varepsilon }_{\varphi }$ and $%
\boldsymbol{\varepsilon }_{\mathrm{d}}=\boldsymbol{\varepsilon }_{\varphi
-\pi /2}$. The stochastic amplitudes associated to these modes satisfy the
Langevin equations 
\begin{subequations}
\begin{align}
\dot{\beta}_{\mathrm{b}}& =\sqrt{2\mathcal{I}}\cos \varphi -\beta _{\mathrm{b%
}}-\Delta \beta _{\mathrm{d}}+\tilde{\beta}_{\mathrm{p}}\beta _{\mathrm{b}%
}^{+}+g\sqrt{\tilde{\beta}_{\mathrm{p}}}\eta _{\mathrm{b}}(\tau ), \\
\dot{\beta}_{\mathrm{b}}^{+}& =\sqrt{2\mathcal{I}}\cos \varphi -\beta _{%
\mathrm{b}}^{+}-\Delta \beta _{\mathrm{d}}^{+}+\tilde{\beta}_{\mathrm{p}%
}^{+}\beta _{\mathrm{b}}+g\sqrt{\tilde{\beta}_{\mathrm{p}}^{+}}\eta _{\mathrm{b%
}}^{+}(\tau ), \\
\dot{\beta}_{\mathrm{d}}& =\sqrt{2\mathcal{I}}\sin \varphi -\beta _{\mathrm{d%
}}+\Delta \beta _{\mathrm{b}}+\tilde{\beta}_{\mathrm{p}}\beta _{\mathrm{d}%
}^{+}+g\sqrt{\tilde{\beta}_{\mathrm{p}}}\eta _{\mathrm{d}}(\tau ), \\
\dot{\beta}_{\mathrm{d}}^{+}& =\sqrt{2\mathcal{I}}\sin \varphi -\beta _{%
\mathrm{d}}^{+}+\Delta \beta _{\mathrm{b}}^{+}+\tilde{\beta}_{\mathrm{p}%
}^{+}\beta _{\mathrm{d}}+g\sqrt{\tilde{\beta}_{\mathrm{p}}^{+}}\eta _{\mathrm{d%
}}^{+}(\tau ),
\end{align}%
where $\tilde{\beta}_{\mathrm{p}}=\sigma -(\beta _{\mathrm{b}}^{2}+\beta _{%
\mathrm{d}}^{2})/2$, $\tilde{\beta}_{\mathrm{p}}^{+}=\sigma -(\beta _{\mathrm{b%
}}^{+2}+\beta _{\mathrm{d}}^{+2})/2$, and 
\end{subequations}
\begin{subequations}
\begin{align}
\eta _{\mathrm{b}}(\tau )& =\frac{1}{\sqrt{2}}[e^{-\mathrm{i}\varphi }\eta
(\tau )+e^{\mathrm{i}\varphi }\eta ^{\ast }(\tau )], \\
\eta _{\mathrm{d}}(\tau )& =\frac{\mathrm{i}}{\sqrt{2}}[e^{-\mathrm{i}%
\varphi }\eta (\tau )-e^{\mathrm{i}\varphi }\eta ^{\ast }(\tau )], \\
\eta _{\mathrm{b}}^{+}(\tau )& =\frac{1}{\sqrt{2}}[e^{\mathrm{i}\varphi
}\eta ^{+}(\tau )+e^{-\mathrm{i}\varphi }\eta ^{+\ast }(\tau )], \\
\eta _{\mathrm{d}}^{+}(\tau )& =-\frac{\mathrm{i}}{\sqrt{2}}[e^{\mathrm{i}%
\varphi }\eta ^{+}(\tau )-e^{-\mathrm{i}\varphi }\eta ^{+\ast }(\tau )],
\end{align}%
behave as real independent white Gaussian noises, that is, defining $\boldsymbol{\eta 
}(\tau )=\text{col}[\eta _{\mathrm{b}}(\tau ),\eta _{\mathrm{b}}^{+}(\tau
),\eta _{\mathrm{d}}(\tau ),\eta _{\mathrm{d}}^{+}(\tau )]$, we have 
\end{subequations}
\begin{equation}
\langle \eta _{j}(\tau )\eta _{l}(\tau ^{\prime })\rangle =\delta
_{jl}\delta (\tau -\tau ^{\prime }).
\end{equation}

Next, we expand the amplitudes as $\beta_{\mathrm{b}}=\sqrt{2I}+b_{\mathrm{b}%
}$, $\beta_{\mathrm{b}}^{+}=\sqrt{2I}+b_{\mathrm{b}}^{+}$, $\beta_{\mathrm{d}%
}=b_{\mathrm{d}}$, and $\beta_{\mathrm{d}}^{+}=b_{\mathrm{d}}^{+}$, and
linearize the equations to first order in the fluctuations and noises,
obtaining the linear system%
\begin{equation}\label{LinLanSym}
\mathbf{\dot{b}}=\mathcal{L}\mathbf{b}+g\sqrt{\sigma-I}\boldsymbol{\eta}%
(\tau),
\end{equation}
where $\mathbf{b}=\text{col}(b_{\mathrm{b}},b_{\mathrm{b}}^{+},b_{\mathrm{d}%
},b_{\mathrm{d}}^{+})$. It is simple to solve analytically this linear
system, and use the solution to evaluate any noise spectrum we want.

\begin{figure}[t]
\begin{center}
\includegraphics[width=\columnwidth]{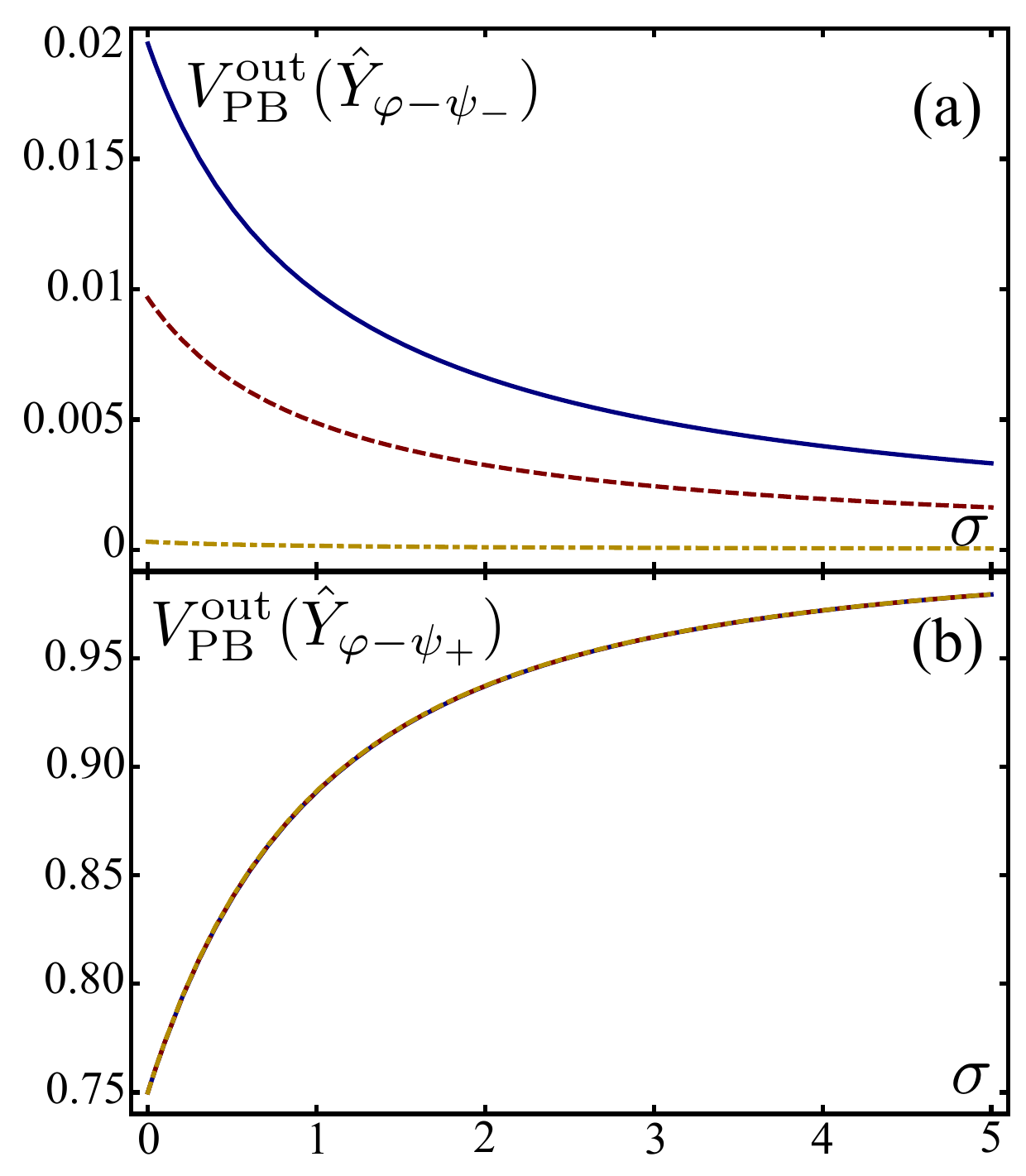}
\end{center}
\caption{Zero-frequency noise spectra of the $\hat{X}$ quadrature of the $\boldsymbol{\varepsilon}_{\varphi-\psi\pm}$\ polarization modes at the Pitchfork bifurcation, for the same values of $\Delta$ as in the previous figure. Note that the mode $\boldsymbol{\varepsilon}_{\varphi-\psi-}$ has large amplitude squeezing for any detuning, while the mode $\boldsymbol{\varepsilon}_{\varphi-\psi_{+}}$ does not have too much squeezing and is basically independent of the detuning.}
\label{fTypeII5}
\end{figure}

\subsection{Entanglement and squeezing at the locking point}

In Appendix \ref{AppendixSym} we solve the linear problem by finding the eigensystem
associated to the linear stability matrix $\mathcal{L}$, arriving to the
following noise spectra for the $\boldsymbol{\varepsilon }_{\varphi \pm \pi
/4}$ polarization modes: 
\begin{subequations}
\label{SimplifiedSpectra}
\begin{align}
V^{\mathrm{out}}(\hat{Y}_{\varphi \pm \pi /4};\Omega )& =1-f_{\pm }(1+\sigma
), \\
V^{\mathrm{out}}(\hat{X}_{\varphi \pm \pi /4};\Omega )& =1+f_{\pm
}(2I+1-\sigma ),
\end{align}%
where 
\end{subequations}
\begin{equation}
f_{\pm }(z)=\frac{4(\sigma -I)[(I\pm \Delta )^{2}+z^{2}+\Omega ^{2}]}{%
(\Delta ^{2}-I^{2}+z^{2})^{2}+2(I^{2}-\Delta ^{2}+z^{2})\Omega ^{2}+\Omega
^{4}}.
\end{equation}

Before analyzing the squeezing levels that can be derived from these
expressions, it is interesting to understand their connection to
entanglement. It is simple to check that the following relations hold: 
\begin{subequations}
\begin{align}
x_{\varphi +\pi /4}& =(x_{\mathrm{s}}^{\varphi +\pi /4}+x_{\mathrm{i}%
}^{-\varphi -\pi /4})/\sqrt{2}, \\
y_{\varphi -\pi /4}& =(x_{\mathrm{s}}^{\varphi +\pi /4}-x_{\mathrm{i}%
}^{-\varphi -\pi /4})/\sqrt{2}, \\
y_{\varphi +\pi /4}& =(y_{\mathrm{s}}^{\varphi +\pi /4}+y_{\mathrm{i}%
}^{-\varphi -\pi /4})/\sqrt{2}, \\
x_{\varphi -\pi /4}& =(y_{\mathrm{i}}^{-\varphi -\pi /4}-y_{\mathrm{s}%
}^{\varphi +\pi /4})/\sqrt{2},
\end{align}%
which show that squeezing in the quadratures of the $\boldsymbol{\varepsilon 
}_{\varphi \pm \pi /4}$ modes imply quantum correlations between the
quadratures of signal and idler. Indeed, whenever the condition 
\end{subequations}
\begin{equation}
V^{\mathrm{out}}\left( \frac{\hat{X}_{\mathrm{s}}^{\varphi _{\mathrm{s}}}-%
\hat{X}_{\mathrm{i}}^{\varphi _{\mathrm{i}}}}{\sqrt{2}};\Omega \right) +V^{%
\mathrm{out}}\left( \frac{\hat{Y}_{\mathrm{s}}^{\varphi _{\mathrm{s}}}+\hat{Y%
}_{\mathrm{i}}^{\varphi _{\mathrm{i}}}}{\sqrt{2}};\Omega \right) <2,
\end{equation}%
is satisfied for some phases $\varphi _{\mathrm{s}}$ and $\varphi _{\mathrm{i%
}}$, it implies that the state of signal and idler is not separable \cite{Duan00,Simon00,NavarreteQICVbook},
which in our case is achieved because the quadratures $\hat{Y}_{\varphi \pm \pi /4}$ are
squeeezed, as we pshow next.

Let us now analyze the entanglement at the locking point. For $1<\sigma
<1+2\Delta $ the Hopf bifurcation is the natural locking point, since it is
the point with which the periodic orbits connect with the stationary solution as the injection parameter $%
\mathcal{I}$ is increased (see Fig. \ref{fTypeII2}b,c); at this point
the zero-frequency noise spectra take the particular form
\begin{subequations}
\begin{align}
& V_{\mathrm{HB}}^{\mathrm{out}}(\hat{Y}_{\varphi \pm \pi /4})=1 \\
& \text{ \ \ \ \ \ \ \ \ }-\frac{8(1+\sigma )[(3+\sigma )^{2}+2(\sigma \pm
\Delta )^{2}+2(1\mp \Delta )^{2}]}{[(3+\sigma )(1+3\sigma )+4\Delta ^{2}]^{2}%
},  \notag \\
& V_{\mathrm{HB}}^{\mathrm{out}}(\hat{X}_{\varphi \pm \pi /4})=\frac{%
(3+\sigma )^{2}+4\Delta ^{2}\pm 4\Delta (\sigma -1)}{(\sigma -1\pm 2\Delta
)^{2}}.
\end{align}%
On the other hand, for $\sigma >1+2\Delta $ the Hopf bifurcation ceases to
exist, and the periodic orbits connect directly with the upper turning point
(see Fig. \ref{fTypeII2}d), in which case the zero-frequency noise spectra read 
\end{subequations}
\begin{subequations}
\begin{align}
V_{\mathrm{+}}^{\mathrm{out}}(\hat{Y}_{\varphi \pm \pi /4})& =1-\frac{%
4(\sigma -I_{+})[(I_{+}\pm \Delta )^{2}+(1+\sigma )^{2}]}{[\Delta
^{2}-I_{+}^{2}+(1+\sigma )^{2}]^{2}}, \\
V_{\mathrm{+}}^{\mathrm{out}}(\hat{X}_{\varphi \pm \pi /4})& =+\infty .
\end{align}%
Note that all these expressions predict squeezing in the $\hat{Y}$ quadratures
of the $\boldsymbol{\varepsilon }_{\varphi \pm \pi /4}$ modes; now, taking
into account that the mean field value of these modes is $\bar{\beta}%
_{\varphi \pm \pi /4}=\sqrt{I}$, this corresponds to phase-squeezing.

In Fig. \ref{fTypeII3} we show the zero-frequency noise spectrum of these
squeezed quadratures $\hat{Y}_{\varphi\pm\pi/4}$ evaluated in the
aforementioned critical points as a function of the pump injection $\sigma$,
and for three different values of the detuning $\Delta$. Note that large
levels of squeezing are obtained in the Hopf bifurcation even when working
up to 20\% above threshold ($\sigma=1.2$).

\subsection{Squeezing at the Pitchfork bifurcation}

Another interesting point is the Pitchfork bifurcation in which the
symmetric solution disappears in favor of another stationary, asymmetric
solution, see Fig. \ref{fTypeII2}. As we already pointed out, in contrast to the injected
two-transverse-mode DOPO \cite{Navarrete11}, we expect perfect squeezing not to appear at this
bifurcation, because the fluctuations of the dark mode are not decoupled
from those of the bright mode below the corresponding threshold. Nevertheless, we prove in this section that large squeezing levels are still attainable.

The first thing to note in this case is that it is more convenient to
analyze the squeezing properties in the polarization basis $\boldsymbol{%
\varepsilon }_{\varphi -\psi _{\pm }}$, where $\psi _{\pm }=\pi /4\pm \arg
\{I-\Delta +\sqrt{I^{2}-\Delta ^{2}}+\mathrm{i}(I-\Delta +\sqrt{I^{2}-\Delta
^{2}})\}$. As shown in Appendix \ref{AppendixSym}, in this basis we get the zero-noise
frequency noise spectra
\end{subequations}
\begin{subequations}\label{VoutPB}
\begin{align}
V_{\mathrm{PB}}^{\mathrm{out}}(\hat{X}_{\varphi -\psi _{-}})& =1-\frac{1}{I_{%
\mathrm{PB}}-\sigma }, \\
V_{\mathrm{PB}}^{\mathrm{out}}(\hat{Y}_{\varphi -\psi _{-}})& =+\infty , \\
V_{\mathrm{PB}}^{\mathrm{out}}(\hat{X}_{\varphi -\psi _{+}})& =1-\frac{I_{%
\mathrm{PB}}-\sigma }{(1+I_{\mathrm{PB}})^{2}}, \\
V_{\mathrm{PB}}^{\mathrm{out}}(\hat{Y}_{\varphi -\psi _{+}})& =1+\frac{I_{%
\mathrm{PB}}-\sigma }{(\sigma +1)^{2}}.
\end{align}
\end{subequations}
In this case the $\hat{Y}$ quadrature of the $\boldsymbol{%
\varepsilon }_{\varphi -\psi _{-}}$ mode is perfectly anti-squeezed; its
complementary, the $\hat{X}$ quadrature of the same mode, is not perfectly
squeezed, but it shows very high noise reduction, as shown in Fig. \ref{fTypeII5}a. On the other hand, the $\boldsymbol{\varepsilon }_{\varphi -\psi _{+}}$
polarization mode shows also noise reduction in its $\hat{X}$ quadrature,
although the squeezing levels are quite modest in this case, see Fig. \ref{fTypeII5}b.

We can understand much better the dependence of these spectra on
the parameters by performing expansions to the leading order in the detuning
(note in particular that the one corresponding to the $\boldsymbol{%
\varepsilon }_{\varphi -\psi _{+}}$ is independent of the detuning):
\begin{subequations}
\begin{align}
V_{\mathrm{PB}}^{\mathrm{out}}(\hat{X}_{\varphi -\psi _{-}})& \approx \frac{%
\Delta ^{2}}{2(\sigma +1)}, \\
V_{\mathrm{PB}}^{\mathrm{out}}(\hat{X}_{\varphi -\psi _{+}})& \approx 1-%
\frac{1}{(\sigma +2)^{2}}.
\end{align}%
Note finally that, in this polarization basis, the steady-state solution reads 
$\bar{\beta}_{\varphi -\psi _{\pm }}=\sqrt{2I_{\mathrm{PB}}}\cos \psi _{\pm }
$, and hence, in both cases we obtain amplitude squeezing, contrary to what
happens at the locking points.

\section{Beyond the symmetric case}

In order to get analytical insight, in the previous sections we have focused
in the case in which signal and idler are detuned symmetrically with respect
to the subharmonic injection at frequency $\omega_0$. In real experiments,
however, it is extremely challenging to meet such a symmetric configuration,
since it requires unfeasible fine-tuning. Hence, in order for our locking
method to be of use, it is important to study whether our predictions persist
when working out of such a symmetric situation, and this is what we prove in
this section. The main difficulty when working out of the symmetric configuration is that
we do not have an analytic solution and stability analysis to rely on, and
hence, we need to resort to numerical tools. Using these, we will show
though that the Hopf instability is still present in the asymmetric case, as
well as large levels of entanglement between signal and idler.

Our starting point is again the normalized equations in which the pump has
been adiabatically eliminated, Eqs. (\ref{FinalLangevin}), but allowing for
general signal and idler detuning, which amounts to replace $\Delta$ by $%
\Delta_\text{s}$ in Eqs. (\ref{NormEqs1}) and (\ref{NormEqs2}), and by $-\Delta_\text{i}$ in Eqs. (\ref{NormEqs3}) and (\ref{NormEqs4}). The first
step consists in finding the classical configuration of the system, what we
do numerically in this case. In particular, we first check that even in this
asymmetric configuration, the classical version of this equations still
possess a Hopf bifurcation above threshold ($\sigma>1$). To this aim, at a
given value of the pump parameter $\sigma$, we start from an injection $%
\mathcal{I}$ large enough so that the system reaches a stationary solution $%
\bar{\beta}_\mathrm{s,i}$, and then decrease the injection gradually until
the real part of one of the eigenvalues of the linear stability matrix gets
as close to zero as we desire, checking that the imaginary part of the
eigenvalue is non-zero. This proves that the Hopf instability is still
present in this asymmetric case, and, moreover, we have checked that if we
keep decreasing the injection, periodic orbits are found as the asymptotic
solution of the system. Hence, again we see that above threshold it is
required a minimum value of the injection to lock the signal and idler
frequencies.

Once we have identified the Hopf bifurcation, which we remind it is the
natural locking point of the system, we compute its quantum properties by
linearizing the Langevin equations, similarly to the symmetric case.
However, in this case we find the eigensystem of the linear stability matrix
numerically for each parameter set. As explained in detail in Appendix \ref{AppendixAsym}, from this
eigensystem we can compute the output field's spectral covariance matrix in the
signal/idler basis, and compute from it the logarithmic negativity
quantifying the entanglement between these two modes following standard
Gaussian techniques \cite{Weedbrook12,NavarreteQICVbook}. We provide all the details in Appendix \ref{AppendixAsym} as well, and
here we just want to compare these levels to the ones obtained in the
symmetric case. In order to do this, we proceed as follows. For every value
of the pump parameter $\sigma$, we choose some distance between the signal
and idler resonances, say $2\Delta>0$. In the symmetric case, this means
that we choose $\Delta_{\mathrm{s}}=-\Delta_{\mathrm{i}}=\Delta $. On the
other hand, as a highly asymmetric case we choose $\Delta_\text{s}=\Delta
+\Delta /2$ and $\Delta _\text{i}=\Delta -\Delta /2$. In Fig. \ref{LogNeg}
we compare the logarithmic negativity obtained in the symmetric (solid line)
and asymmetric (markers) cases for the three values of $\Delta$ that we also
chose in Figs. \ref{fTypeII3} and \ref{fTypeII5}. Remarkably, we can see that, not only the entanglement levels are
also high in the asymmetric case, but they coincide within the numerical accuracy with the
ones of the symmetric case. This suggests that the entanglement properties
of the system depend only on the distance between the signal and idler
resonances, and not on how they are disposed with respect to the frequency
of the subharmonic injection, which is a most important conclusion for
experiments.

\begin{figure}[t]
\centering
\includegraphics[width=.90\columnwidth]{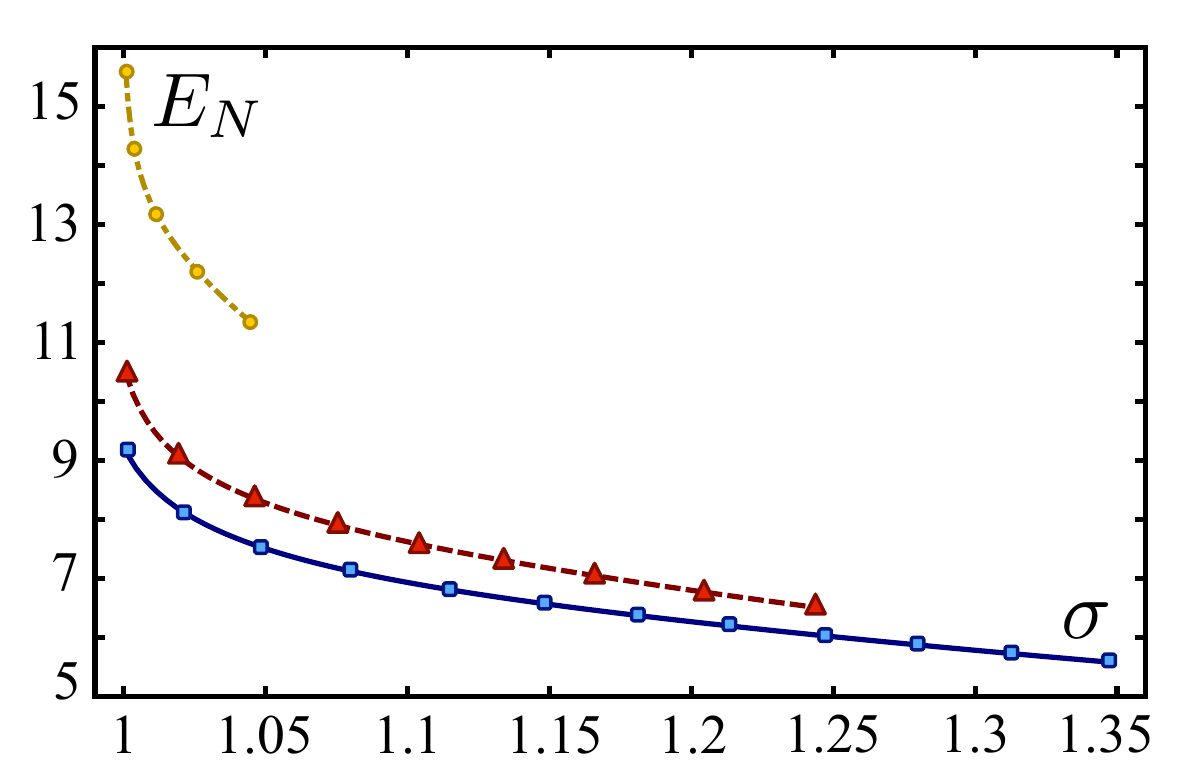}
\caption{Logarithmic negativity ($E_{N}$) as a function of the pump
parameter $\protect\sigma$ at the Hopf bifurcation, which corresponds to the
minimum value of the injection $\mathcal{I}$ for which the oscillation
frequencies of signal and idler get locked. The solid curves correspond to
the analytical solution that we found for the symmetric case within the
linearized theory, while the markers are found numerically for the
asymmetric case as explained in the text. Three values of $\Delta$ have been
chosen, coinciding with the ones in the previous figures: 0.2 (solid blue, squares), 0.14 (dashed red, triangles) and 0.025 (dashed-dotted yellow, circles).}
\label{LogNeg}
\end{figure}

\section{Conclusions}

In conclusion, in this work we have put forward a method to obtain exact frequency degeneracy in type II OPOs, which is based on the injection of a laser field at half the frequency of the pump laser. We have studied the impact that such subharmonic injection has on the entanglement generated on the down-converted fields, proving that large quantum correlations are still present at the locking region. Hence, this technique offers an easily tunable alternative to more invasive techniques which require the introduction of additional optical elements in the cavity. Apart from large levels of entanglement at the locking bifurcation, we have also identified an additional (static) instability where the polarization mode orthogonal to the classically excited one is in a highly squeezed vacuum state.

Let us finally note that we have also analyzed the case in which the subharmonic injection is not in phase with the pump beam (amplification regime), but is phase-shifted by $\pi/2$ (attenuation regime), finding similar results that will be shown elsewhere.

\begin{acknowledgements}
We have benefited from discussions with Claude Fabre, Ra\'ul Garc\'ia-Patr\'on, Talitha Weiss, Andrea Mari, and G\'eza Giedke. This work has been supported by the Spanish Government and the European Union FEDER through Projects FIS2011-26960 and FIS2014-60715-P. C.N.-B. acknowledges support from the from the ERC OPTOMECH and the Alexander von Humboldt foundation through its fellowship for postdoctoral researchers.
\end{acknowledgements}

\appendix

\section{Manipulating the linearized Langevin equations}\label{Appendix}

Our analysis of the quantum properties of the down-converted field was based
on the linearized Langevin equations. In this appendix we will show
explicitly how we have dealt with these equations in order to obtain the quantities of interest, both in the symmetric and asymmetric cases. Conceptually, the approach we use is the same in the symmetric and asymmetric cases:
we solve the linear system by making use of the eigensystem of the stability
matrix. However, in the symmetric case we will be able to find the eigensystem analytically, while in the asymmetric case only numerically. Let us then start by commenting on some general aspects, and then
particularize to our problems at hand.

In general, the linearized Langevin equations can be written in the form 
\end{subequations}
\begin{equation}
\mathbf{\dot{b}}=\mathcal{L}\mathbf{b}+g\sqrt{|\bar{\beta}_{\mathrm{p}}|}%
\boldsymbol{\eta }(\tau ).  \label{GenLinLanApp}
\end{equation}
In this expression $\bar{\beta}_{\mathrm{p}}=\sigma -\bar{\beta}_{\mathrm{s}}\bar{\beta}_{%
\mathrm{i}}$, $\mathbf{b}$ is a vector containing the quantum fluctuations
of the stochastic amplitudes in the polarization basis that we choose to
write the equations on, $\mathcal{L}$ is the corresponding linear stability
matrix, and we assume that the components of the noise vector obey the
two-time correlators $\langle \eta _{m}(\tau )\eta _{n}(\tau ^{\prime
})\rangle =\mathcal{S}_{mn}\delta (\tau -\tau ^{\prime })$, with $\mathcal{S}
$ some matrix.

Given this equation, we proceed by finding the left eigenvectors $\{\mathbf{u%
}_{j}\}_{j=1,2,3,4}$ defined by $\mathbf{u}_{j}^{\dagger }\mathcal{L}%
=\lambda _{j}\mathbf{u}_{j}^{\dagger }$ or, equivalently, $\mathcal{L}^{\dagger}%
\mathbf{u}_{j}=\lambda _{j}^{\ast }\mathbf{u}_{j}$ (note that they are
defined as column vectors). The corresponding eigenvalues are denoted by $%
\lambda _{j}$. Acting on Eq. (\ref{GenLinLanApp}) with $\mathbf{u}%
_{j}^{\dagger }$ on the left, and defining the projections $c_{j}(\tau )=%
\mathbf{u}_{j}^{\dagger }\mathbf{b}(\tau )$, we obtain 
\begin{equation}
\dot{c}_{j}=\lambda _{j}c_{j}+g\sqrt{|\bar{\beta}_{\mathrm{p}}|}\mathbf{u}%
_{j}^{\dagger }\boldsymbol{\eta }(\tau ),
\end{equation}%
which has the asymptotic ($\tau \gg -\text{Re}\{\lambda _{j}\}^{-1}$
$\forall j$) solution
\begin{equation}
c_{j}(\tau )=g\sqrt{|\bar{\beta}_{\mathrm{p}}|}\int_{0}^{\tau }d\tau
e^{\lambda _{j}(\tau -\tau ^{\prime })}\mathbf{u}_{j}^{\dagger }\boldsymbol{%
\eta }(\tau ^{\prime }),
\end{equation}%
leading to the asymptotic correlation functions%
\begin{equation}
\langle c_{j}(\tau )c_{l}(\tau ^{\prime })\rangle =-\frac{g^{2}|\bar{\beta}_{%
\mathrm{p}}|\mathbf{u}_{j}^{\dagger }\mathcal{S}\mathbf{u}_{l}^{\ast }}{%
\lambda _{j}+\lambda _{l}}\times\left\{ 
\begin{array}{cc}
e^{\lambda _{l}(\tau ^{\prime }-\tau)} & \tau ^{\prime }>\tau \\ 
e^{\lambda _{j}(\tau -\tau ^{\prime })} & \tau ^{\prime }<\tau%
\end{array}%
\right. \hspace{-1.5mm},
\end{equation}%
and ultimately to the asymptotic spectra%
\begin{eqnarray}
\mathcal{C}_{jl}(\Omega ) &=& \lim_{\tau\rightarrow\infty}\int_{-\infty }^{+\infty }d\tau ^{\prime }e^{-%
\mathrm{i}\Omega \tau'}\langle c_{j}(\tau )c_{l}(\tau +\tau ^{\prime
})\rangle  \notag
\\
&=&\frac{g^{2}|\bar{\beta}_{\mathrm{p}}|\mathbf{u}_{l}^{\dagger }\mathcal{S}%
\mathbf{u}_{j}^{\ast }}{(\lambda _{j}+\mathrm{i}\Omega )(\lambda _{l}-%
\mathrm{i}\Omega )},
\end{eqnarray}%
which define a matrix $\mathcal{C}$. The noise spectrum (\ref{Vout}) of any
quadrature, or even more complicated objects such as the spectral covariance
matrix in any polarization basis, can be evaluated by making a proper
combination of these spectra, as we will see shortly.

\subsection{Symmetric configuration}\label{AppendixSym}

In the case of the symmetric configuration, the linearized Langevin
equations take the form (\ref{LinLanSym}) in the bright/dark basis, leading to $|\bar{%
\beta}_{\mathrm{p}}|=\sigma -I$, $\mathcal{S}=\mathbb{1}$, where $\mathbb{1}$ is the identity matrix, and a linear stability matrix $\mathcal{L}$ given
by Eq. (\ref{Lbd}). In order to apply the general expressions above in this
symmetric configuration, it is convenient to analyze separately the cases $%
I<\Delta $ and $I>\Delta $, since $\mathcal{L}$ becomes singular at $%
I=\Delta $.

\subsubsection{Eigensystem and noise spectra for $I<\Delta $}

In the $I<\Delta $ case, the eigenvalues read 
\begin{subequations}
\begin{align}
\lambda _{1}& =-1-\sigma -\mathrm{i}\sqrt{\Delta ^{2}-I^{2}}, \\
\lambda _{2}& =-1-\sigma +\mathrm{i}\sqrt{\Delta ^{2}-I^{2}}, \\
\lambda _{3}& =-1+\sigma -2I-\mathrm{i}\sqrt{\Delta ^{2}-I^{2}}, \\
\lambda _{4}& =-1+\sigma -2I+\mathrm{i}\sqrt{\Delta ^{2}-I^{2}},
\end{align}%
with corresponding left eigenvectors 
\end{subequations}
\begin{subequations}
\begin{align}
\mathbf{u}_{1}& =\text{col}(e^{-\mathrm{i}\phi /2},-e^{-\mathrm{i}\phi
/2},e^{\mathrm{i}\phi /2},-e^{\mathrm{i}\phi /2}), \\
\mathbf{u}_{2}& =\text{col}(e^{\mathrm{i}\phi /2},-e^{\mathrm{i}\phi /2},e^{-%
\mathrm{i}\phi /2},-e^{-\mathrm{i}\phi /2}), \\
\mathbf{u}_{3}& =\text{col}(e^{-\mathrm{i}\phi /2},e^{-\mathrm{i}\phi /2},e^{%
\mathrm{i}\phi /2},e^{\mathrm{i}\phi /2}), \\
\mathbf{u}_{4}& =\text{col}(e^{\mathrm{i}\phi /2},e^{\mathrm{i}\phi /2},e^{-%
\mathrm{i}\phi /2},e^{-\mathrm{i}\phi /2}),
\end{align}%
where $\phi =\arg \{I+\mathrm{i}\sqrt{\Delta ^{2}-I^{2}}\}$.

Since the eigenvalues are complex, it is clear that the projections $c_{j}$
cannot be directly proportional to observable quantities. However, one can easily show
that simple combinations of them are indeed proportional to the quadratures
of the $\boldsymbol{\varepsilon }_{\varphi \pm \pi /4}$ polarization modes: 
\end{subequations}
\begin{subequations}
\begin{align}
c_{1}+c_{2}& =2\mathrm{i}\sqrt{1+I/\Delta }\delta y_{\varphi -\pi /4}, \\
c_{1}-c_{2}& =-2\sqrt{1-I/\Delta }\delta y_{\varphi +\pi /4}, \\
c_{3}+c_{4}& =2\sqrt{1+I/\Delta }\delta x_{\varphi -\pi /4}, \\
c_{3}-c_{4}& =2\mathrm{i}\sqrt{1-I/\Delta }\delta x_{\varphi +\pi /4}.
\end{align}%
Hence, we get the noise spectra 
\end{subequations}
\begin{subequations}
\label{Spectra2}
\begin{align}
V^{\mathrm{out}}(\hat{Y}_{\varphi -\pi /4};\Omega )& =1-[2g^{2}(1+I/\Delta )]^{-1} \\
& \hspace{-5mm}\times \lbrack \mathcal{C}_{11}(\Omega )+\mathcal{C}_{22}(\Omega
)+\mathcal{C}_{21}(\Omega )+\mathcal{C}_{12}(\Omega )],  \notag \\
V^{\mathrm{out}}(\hat{Y}_{\varphi +\pi /4};\Omega )& =1+[2g^{2}(1-I/\Delta )]^{-1} \\
& \hspace{-5mm}\times \lbrack \mathcal{C}_{11}(\Omega )+\mathcal{C}_{22}(\Omega
)-\mathcal{C}_{21}(\Omega )-\mathcal{C}_{12}(\Omega )],  \notag \\
V^{\mathrm{out}}(\hat{X}_{\varphi -\pi /4};\Omega )& =1+[2g^{2}(1+I/\Delta )]^{-1} \\
& \hspace{-5mm}\times \lbrack \mathcal{C}_{33}(\Omega )+\mathcal{C}_{44}(\Omega
)+\mathcal{C}_{34}(\Omega )+\mathcal{C}_{43}(\Omega )],  \notag \\
V^{\mathrm{out}}(\hat{X}_{\varphi +\pi /4};\Omega )& =1-[2g^{2}(1-I/\Delta )]^{-1} \\
& \hspace{-5mm}\times \lbrack \mathcal{C}_{33}(\Omega )+\mathcal{C}_{44}(\Omega
)-\mathcal{C}_{34}(\Omega )-\mathcal{C}_{43}(\Omega )].  \notag
\end{align}%
These spectra have actually fairly simple analytical expressions in terms of
the system parameters, expressions that we gave explicitly in Eqs. (\ref{SimplifiedSpectra}) in
the main text.

\subsubsection{Eigensystem and noise spectra for $I>\Delta $}

In the $I>\Delta $ case, defining the functions $F_{\pm }=I\pm \sqrt{%
I^{2}-\Delta ^{2}}>0$, the left eigensystem of $\mathcal{L}$ is easily found
to be 
\end{subequations}
\begin{subequations}
\begin{align}
\mathbf{u}_{1}& =\text{col}(F_{+},-F_{+},\Delta ,-\Delta ), \\
\mathbf{u}_{2}& =\text{col}(F_{-},-F_{-},\Delta ,-\Delta ), \\
\mathbf{u}_{3}& =\text{col}(F_{+},F_{+},\Delta ,\Delta ), \\
\mathbf{u}_{4}& =\text{col}(F_{-},F_{-},\Delta ,\Delta ),
\end{align}%
with corresponding eigenvalues 
\end{subequations}
\begin{subequations}
\begin{align}
\lambda _{1}& =-1-\sigma -\sqrt{I^{2}-\Delta ^{2}}, \\
\lambda _{2}& =-1-\sigma +\sqrt{I^{2}-\Delta ^{2}}, \\
\lambda _{3}& =-1+\sigma -2I-\sqrt{I^{2}-\Delta ^{2}}, \\
\lambda _{4}& =-1+\sigma -2I+\sqrt{I^{2}-\Delta ^{2}}.
\end{align}

Let us define the amplitude and phase of $F_{\pm }+\mathrm{i}\Delta $ as $%
M_{\pm }$ and $\psi _{\pm }$ respectively, which can be written as 
\end{subequations}
\begin{subequations}
\begin{align}
M_{\pm }& =2I\left( I\pm \sqrt{I^{2}-\Delta ^{2}}\right) , \\
\psi _{\pm }& =\frac{\pi }{4}\mp \psi ,
\end{align}%
with $\psi =\arg \{I-\Delta +\sqrt{I^{2}-\Delta ^{2}}+\mathrm{i}(I-\Delta +%
\sqrt{I^{2}-\Delta ^{2}})\}$. In this case the eigenvalues are real, and it
is therefore possible to find a relation between the projections and the
quadratures of modes with polarization $\boldsymbol{\varepsilon }_{\varphi-\psi_\pm}$: 
\end{subequations}
\begin{subequations}
\begin{align}
c_{1}& =\sqrt{M_{+}}\mathrm{i}\delta y_{\varphi -\psi _{+}}, \\
c_{2}& =\sqrt{M_{-}}\mathrm{i}\delta y_{\varphi -\psi _{-}}, \\
c_{3}& =\sqrt{M_{+}}\delta x_{\varphi -\psi _{+}}, \\
c_{4}& =\sqrt{M_{-}}\delta x_{\varphi -\psi _{-}},
\end{align}%
leading to the noise spectra 
\end{subequations}
\begin{subequations}
\begin{align}
V^{\mathrm{out}}(\hat{Y}_{\varphi -\psi _{+}};\Omega )& =1-\frac{2}{%
g^{2}M_{+}}\mathcal{C}_{11}(\Omega ) \\
& =1-\frac{4(\sigma -I)}{(1+\sigma +\sqrt{I^{2}-\Delta ^{2}})^{2}+\Omega ^{2}%
},  \notag \\
V^{\mathrm{out}}(\hat{Y}_{\varphi -\psi _{-}};\Omega )& =1-\frac{2}{%
g^{2}M_{-}}\mathcal{C}_{22}(\Omega ) \\
& =1-\frac{4(\sigma -I)}{(1+\sigma -\sqrt{I^{2}-\Delta ^{2}})^{2}+\Omega ^{2}%
},  \notag \\
V^{\mathrm{out}}(\hat{X}_{\varphi -\psi _{+}};\Omega )& =1+\frac{2}{%
g^{2}M_{+}}\mathcal{C}_{33}(\Omega ) \\
& =1+\frac{4(\sigma -I)}{(1-\sigma +2I+\sqrt{I^{2}-\Delta ^{2}})^{2}+\Omega
^{2}},  \notag \\
V^{\mathrm{out}}(\hat{X}_{\varphi -\psi _{-}};\Omega )& =1+\frac{2}{%
g^{2}M_{-}}\mathcal{C}_{44}(\Omega ) \\
& =1+\frac{4(\sigma -I)}{(1-\sigma +2I-\sqrt{I^{2}-\Delta ^{2}})^{2}+\Omega
^{2}}.  \notag
\end{align}%
This expressions, particularized to $\Omega =0$ and the pitchfork
bifurcation $I=I_\text{PB}$ are the ones we gave in Eqs. (\ref{VoutPB}).

In order to compare with the $I<\Delta $ case, it is also convenient to
analyze the noise spectra in the $\boldsymbol{\varepsilon }_{\varphi \pm \pi
/4}$ polarization basis. For this, we now relate these mode's quadratures
modes with the projections $c_{j}$. In particular, it is easy to find 
\end{subequations}
\begin{subequations}
\begin{align}
\frac{c_{1}}{\sqrt{M_{+}}}+\frac{c_{2}}{\sqrt{M_{-}}}& =\sqrt{2\left(1+\frac{\Delta}{I}\right)}%
\mathrm{i}\delta y_{\varphi -\pi /4}, \\
\frac{c_{1}}{\sqrt{M_{+}}}-\frac{c_{2}}{\sqrt{M_{-}}}& =\sqrt{2\left(1-\frac{\Delta}{I}\right)}%
\mathrm{i}\delta y_{\varphi +\pi /4}, \\
\frac{c_{3}}{\sqrt{M_{+}}}+\frac{c_{4}}{\sqrt{M_{-}}}& =\sqrt{2\left(1+\frac{\Delta}{I}\right)}%
\delta x_{\varphi -\pi /4}, \\
\frac{c_{3}}{\sqrt{M_{+}}}-\frac{c_{4}}{\sqrt{M_{-}}}& =\sqrt{2\left(1-\frac{\Delta}{I}\right)}%
\delta x_{\varphi +\pi /4}.
\end{align}%
where the identities $\sqrt{2}\cos \psi =\sqrt{1+\Delta /I}$ and $\sqrt{2}%
\sin \psi =\sqrt{1-\Delta /I}$ are useful when checking this relations.
Hence, the noise spectrum of the corresponding quadratures can be written as 
\end{subequations}
\begin{subequations}
\label{Spectra1}
\begin{align}
V^{\mathrm{out}}(\hat{Y}_{\varphi -\pi /4};\Omega )& =1-\left[g^{2}\left(1+\frac{\Delta}{I}\right)\right]^{-1} \\
& \hspace{-1.3cm}\times \left[ \frac{\mathcal{C}_{11}(\Omega )}{M_{+}}+\frac{%
\mathcal{C}_{22}(\Omega )}{M_{-}}+\frac{\mathcal{C}_{21}(\Omega )+\mathcal{C}%
_{12}(\Omega )}{\sqrt{M_{+}M_{-}}}\right],  \notag \\
V^{\mathrm{out}}(\hat{Y}_{\varphi +\pi /4};\Omega )& =1-\left[g^{2}\left(1-\frac{\Delta}{I}\right)\right]^{-1} \\
& \hspace{-1.3cm}\times \left[ \frac{\mathcal{C}_{11}(\Omega )}{M_{+}}+\frac{%
\mathcal{C}_{22}(\Omega )}{M_{-}}-\frac{\mathcal{C}_{21}(\Omega )+\mathcal{C}%
_{12}(\Omega )}{\sqrt{M_{+}M_{-}}}\right] ,  \notag \\
V^{\mathrm{out}}(\hat{X}_{\varphi -\pi /4};\Omega )& =1+\left[g^{2}\left(1+\frac{\Delta}{I}\right)\right]^{-1} \\
& \hspace{-1.3cm}\times \left[ \frac{\mathcal{C}_{33}(\Omega )}{M_{+}}+\frac{%
\mathcal{C}_{44}(\Omega )}{M_{-}}+\frac{\mathcal{C}_{34}(\Omega )+\mathcal{C}%
_{43}(\Omega )}{\sqrt{M_{+}M_{-}}}\right],  \notag \\
V^{\mathrm{out}}(\hat{X}_{\varphi +\pi /4};\Omega )& =1+\left[g^{2}\left(1-\frac{\Delta}{I}\right)\right]^{-1}\\
& \hspace{-1.3cm}\times \left[ \frac{\mathcal{C}_{33}(\Omega )}{M_{+}}+\frac{%
\mathcal{C}_{44}(\Omega )}{M_{-}}-\frac{\mathcal{C}_{34}(\Omega )+\mathcal{C}%
_{43}(\Omega )}{\sqrt{M_{+}M_{-}}}\right].  \notag
\end{align}%
It is again easy to check that, in terms of the system parameters, these
combinations read as given in Eqs. (\ref{SimplifiedSpectra}), and hence they coincide with the
expressions found in the $I>\Delta $ case. This means that, even though the
eigensystems are very different in the $I>\Delta $ and $I<\Delta $ cases,
and furthermore the matrix $\mathcal{L}$ cannot be diagonalized in the $%
I=\Delta $ limit, this mathematical pathology\ is not present in the
physical observables. This is indeed characteristic of detuned nonlinear
quantum-optical cavities.

\subsection{Asymmetric configuration}\label{AppendixAsym}

In the case of the asymmetric configuration, we work in the signal/idler
basis, where we find the classical solution $\bar{\beta}_{\mathrm{s,i}}$
numerically as explained in the text for each choice of parameters. In this
case, we then have $\mathbf{b}=\text{col}(b_{\mathrm{s}},b_{\mathrm{s}%
}^{+},b_{\mathrm{i}},b_{\mathrm{i}}^{+})$, 
\end{subequations}
\begin{equation}
\mathcal{L}=\left( 
\begin{matrix}
\Theta_\text{s} & 0 & -\bar{\beta}_\text{s}\bar{\beta}_\text{i}^* & \sigma \\ 
0 & \Theta_\text{s}^* & \sigma & -\bar{\beta}_\text{s}^*\bar{\beta}_\text{i} \\ 
-\bar{\beta}_\text{s}^*\bar{\beta}_\text{i} & \sigma & \Theta_\text{i} & 0 \\ 
\sigma & -\bar{\beta}_\text{s}\bar{\beta}_\text{i}^* & 0 & \Theta_\text{i}^*%
\end{matrix}%
\right) ,
\end{equation}
with $\Theta_\text{s,i}=-1-\text{i}\Delta_\text{s,i}-|\bar{\beta}_\text{i,s}|^2$, and
\begin{equation}
\mathcal{S}=\left( 
\begin{matrix}
0 & 0 & e^{2\mathrm{i}\varphi _{\mathrm{p}}} & 0 \\ 
0 & 0 & 0 & e^{-2\mathrm{i}\varphi _{\mathrm{p}}} \\ 
e^{2\mathrm{i}\varphi _{\mathrm{p}}} & 0 & 0 & 0 \\ 
0 & e^{-2\mathrm{i}\varphi _{\mathrm{p}}} & 0 & 0%
\end{matrix}%
\right) ,
\end{equation}%
with $\varphi_{\mathrm{p}}=\mathrm{arg}\{\bar{\beta}_\text{p}\}$. We find
the eigensystem of $\mathcal{L}$ numerically for each
choice of the system parameters.

We can characterize the quantum state of the output field by the spectral
covariance matrix. Collecting the normalized stochastic quadratures of
signal and idler in a vector $\mathbf{r}=(x_{\mathrm{s}},y_{\mathrm{s}},x_{%
\mathrm{i}},y_{\mathrm{i}})$, this can be evaluated as 
\begin{equation}
\mathcal{V}(\Omega )=\mathbb{1}+\frac{2}{g^{2}}\int_{-\infty }^{+\infty
}d\tau \mathcal{M}(\tau )e^{-\mathrm{i}\Omega \tau },  \label{2Mcovarianze}
\end{equation}%
where the elements of the normally-ordered two-time correlation matrix $%
\mathcal{M}$ are given by 
\begin{equation}
\mathcal{M}_{jl}(\tau)=\hspace{-0.5mm}\lim_{\tau ^{\prime }\rightarrow
\infty }\frac{\langle \delta r_{j}(\tau ^{\prime })\delta r_{l}(\tau -\tau
^{\prime })\hspace{-0.5mm}+\hspace{-0.5mm}\delta r_{l}(\tau ^{\prime
})\delta r_{j}(\tau -\tau ^{\prime })\rangle }{2}.
\end{equation}
At the end of this section we explain how this two-mode covariance matrix
allows for a characterization of the entanglement between the signal and
idler modes. But before that, let us show how we can compute it from the
solution that we found for the linearized problem, in particular from the
spectral correlation matrix $\mathcal{C}(\Omega )$ of the projections. Note
that the relation between the quadrature fluctuations $\delta \mathbf{r}$
and the quantum fluctuations $\mathbf{b}$ can be written in matrix form as $%
\delta \mathbf{r}(\tau)=\mathcal{R}\mathbf{b}(\tau)$ with
\begin{equation}
\mathcal{R}=\left( 
\begin{matrix}1 & 1 \\ -i & i\end{matrix}\right)\oplus\left( 
\begin{matrix}1 & 1 \\ -i & i\end{matrix}\right),
\end{equation}
while defining the vector of projections $\mathbf{c}=\mathrm{col}%
(c_{1},c_{2},c_{3},c_{4})$ and the matrix of left-eigenvectors $\mathcal{U}=%
\mathrm{col}(\mathbf{u}_{1}^{\dagger },\mathbf{u}_{2}^{\dagger },\mathbf{u}%
_{3}^{\dagger },\mathbf{u}_{4}^{\dagger })$, we can write $\mathbf{b}(\tau)=%
\mathcal{U}^{-1}\mathbf{c}(\tau)$. Hence, we see that we can write the
quadrature-vector in terms of the projection-vector as $\delta \mathbf{r}(\tau)=%
\mathcal{R}\mathcal{U}^{-1}\mathbf{c}(\tau)$, leading to%
\begin{equation}
\mathcal{V}(\Omega )=\mathbb{1}+\frac{1}{g^{2}}\mathcal{R}\mathcal{U}^{-1}[%
\mathcal{C}(\Omega )+\mathcal{C}^{T}(\Omega )]\mathcal{U}^{-1T}\mathcal{R}%
^{T}.
\end{equation}%
Let us remark that this expression can be efficiently evaluated numerically
once we have identified the classical stationary solution at the Hopf
bifurcation, $\bar{\beta}_{\mathrm{s,i}}$, from which we derive the linear
stability matrix $\mathcal{L}$, its eigensystem, and from it $\mathcal{U}%
^{-1}$ as well as the spectral correlation matrix $\mathcal{C}(\Omega )$. In
the following we take $\Omega =0$ as this is the value of the noise
frequency that leads to the largest levels of entanglement in the symmetric
case.

Having the covariance matrix, we are now ready to analyze the entanglement
between the signal and idler modes. In order to be numerically efficient, we
choose to quantify the entanglement between these two modes via the
logarithmic negativity, which is an entanglement monotone, albeit not a proper measure
\cite{NavarreteQICVbook}. Since by construction, the linearized approach generates a Gaussian state for the system, the logarithmic negativity can be easily computed from the two-mode spectral covariance matrix by following standard techniques, see for example \cite{Weedbrook12,NavarreteQICVbook}. In particular, defining the partially-transposed spectral covariance matrix
\begin{equation}
\tilde{\mathcal{V}}=\mathcal{Z}\mathcal{V}(0)\mathcal{Z}\equiv
\left(\begin{array}{cc}
A & C
\\ 
C^T & B
\end{array}\right),
\end{equation}
where $\mathcal{Z}=\text{diag}(1,1,1,-1)$, the logarithmic negativity takes the expression
\begin{equation}
E_N=-\sum_{j=\pm}
\left\{\begin{array}{cc}
\log\tilde{\nu}_j & \tilde{\nu}_j<1
\\ 
0 & \tilde{\nu}_j\geq 1
\end{array}\right.,	
\end{equation}
where $\tilde{\nu}_\pm$ are the symplectic eigenvalues associated to $\tilde{\mathcal{V}}$, which can be found from
\begin{equation}
\tilde{\nu}_\pm^2=\frac{\tilde{\Delta}\pm\sqrt{\tilde{\Delta}^2-4\det\tilde{\mathcal{V}}}}{2},
\end{equation}
with $\tilde{\Delta}=\det A+\det B+2 \det C$.

\end{document}